\documentclass[aps,prmaterials,twocolumn,superscriptaddress,amsmath,amssymb,showpacs,english]{revtex4-2}

\usepackage{graphicx}
\usepackage{epstopdf}
\usepackage{dcolumn}
\usepackage{bm}
\usepackage{color} 
\usepackage{ulem}
\usepackage{xspace}
\usepackage{multirow}
\usepackage{hyperref}
\usepackage{physics}
\usepackage{here}
\usepackage[caption=false]{subfig}

\newcommand{\Riken}{RIKEN Center for Emergent Matter Science, 2-1 Hirosawa, Wako, 351-0198, Japan}
\newcommand{\UniRoma}{Dipartimento di Fisica, Universit\`a di Roma La Sapienza, Piazzale Aldo Moro 5, I-00185 Roma, Italy}
\newcommand{\UniTokyoPhy}{Department of Applied Physics, University of Tokyo, Tokyo 113-8656, Japan}
\newcommand{\UniTokyoFron}{Graduate School of Frontier Science, University of Tokyo, Kashiwa, Chiba 277-0882, Japan}

\makeatletter
\newcommand*{\rom}[1]{\expandafter\@slowromancap\romannumeral #1@}
\makeatother

\definecolor{green}{rgb}{0,0.6,0.1}

\begin{document}
	
\title{First-Principles Design of Halide-Reduced Electrides: \\ Magnetism and Topological Phases}

\author{Tonghua Yu}             \email{yutonghua@g.ecc.u-tokyo.ac.jp} \affiliation{\UniTokyoPhy}  
\author{Motoaki Hirayama}       \affiliation{\UniTokyoPhy} \affiliation{\Riken}
\author{Jos\'e A. Flores-Livas} \affiliation{\Riken}       \affiliation{\UniRoma}
\author{Marie-Therese Huebsch}  \affiliation{\Riken}       \affiliation{\UniTokyoFron}
\author{Takuya Nomoto}          \affiliation{\UniTokyoPhy}
\author{Ryotaro Arita}          \affiliation{\UniTokyoPhy} \affiliation{\Riken}

\begin{abstract}
We propose a design scheme for potential electrides derived from conventional materials. 
Starting with rare-earth-based ternary halides, we exclude halogens and perform global structure optimization to obtain thermodynamically stable or metastable phases but having an excess of electrons confined inside interstitial cavities. Then, spin-polarized interstitial states are induced by chemical substitution with magnetic lanthanides. 
To demonstrate the capability of our approach, we test with 11 ternary halides and successfully predict 30 stable and metastable phases of nonmagnetic electrides subject to 3 different stoichiometric categories, and successively 28 magnetic electrides via chemical substitution with Gd. 56 out of these 58 designed electrides are discovered for the first time. 
Two electride systems, the monoclinic $A$C ($A=$ La, Gd) and the orthorhombic $A_2$Ge ($A=$ Y, Gd), are thoroughly studied to exemplify the set of predicted crystals. Interestingly, both systems turn out to be topological nodal line electrides (TNLE) in the absence of spin-orbit coupling and manifest spin-polarized interstitial states in the case of $A=$ Gd. 
Our work establishes a novel computational approach of functional electrides design and highlights the magnetism and topological phases embedded in electrides.
\end{abstract}

\maketitle

\section{Introduction} 

Electrides are a novel class of crystals with vacant cavities accommodating intrinsic excess electrons, which are not tightly constrained by certain nuclei, but on the contrary, form ionic bonds with the cationic framework, playing the role of anions \cite{dye1990663, dye2003electrons}. In contrast to random defects or color centers, sites of interstitial electrons in electrides adhere to the translational periodicity. Since the experimental synthesis of the first room-temperature stable inorganic electride [Ca$_{24}$Al$_{28}$O$_{64}$]$^{4+}(4e^-)$ (C12A7:2e$^-$) \cite{matsuishi2003high}, electrides have attracted great attention among the communities of materials science, chemistry and condensed matter physics, owing to the intriguing properties exhibited by non-nucleus-bound electrons. For instance, in magnetic electrides the spin-unpaired anionic electrons are well suited for spin injection in a spintronic device \cite{sui2019prediction, wolf2001spintronics}, on account of their low work function. In addition, the interstitial electronic states are found to favor band inversion near the Fermi level, giving rise to the high tendency for electrides towards topological phases \cite{hirayama2018electrides, zhu2019computational}. One could therefore implement electrides as an excellent testbed to study topological physics \cite{PhysRevLett.120.026401, huang2018topological, PhysRevLett.123.206402, zhang2019topological, nie2020application}.

Hitherto, known electrides especially magnetic electrides, however, are extremely scarce\cite{liu2020electrides}. Consecutive efforts have been devoted by a number of researchers to the computational discovery of new electrides. Approaches of the mainstream involve: structure screening from the materials database \cite{zhu2019computational, PhysRevX.4.031023, Tada2014High-Throughput, Burton2018High-Throughput}, crystal structure prediction (CSP) for unknown prototype structures \cite{tsuji2016structural, ming2016first, zhang2017computer, wang2017exploration}, and functionality-tailored electrides design \cite{PhysRevLett.120.026401, PhysRevMaterials.3.024205}. Here we present an alternative computational strategy to design new electrides. It is widely noticed that every natural electride corresponds to a potential oxidized form, such as Ca$_2$N to Ca$_2$NCl, Y$_2$C to Y$_2$H$_2$C. One is therefore motivated to ask whether the reverse direction holds \cite{liu2020electrides}. In other words, if the anions with large electronegativity is taken away from a selected compound (e.g. remove Cl from Ca$_2$NCl), the remaining structure (Ca$_2$N) is suspected to be a possible electride.
To ascertain this question, we mainly focus on ternary halides containing trivalent rare-earth elements, i.e. scandium, yttrium, and lanthanides. Optimizing the reduced structure using a CSP method, a stable or metastable crystal is expected to be the candidate of a binary electride. Chemical substitution of magnetic lanthanides is followed to induce spin-polarized interstitial electronic states, by which magnetic electrides can be constructed. Starting from 11 halides serving as structure hosts, our method predicts 30 thermodynamically stable and metastable nonmagnetic electrides, 29 of which are newly designed. While retaining the thermodynamic (meta)stability, we replace cationic elements with the lanthanide Gd to induce spin polarization for these crystals. This chemical substitution procedure successfully generates 28 candidate structures of magnetic electrides, with 27 of them proposed for the first time. Electronic properties of two representative systems are analyzed in depth. In addition to electride and magnetism features, topological nodal lines in the electronic bands of these two systems are unveiled as well, further confirming the intimate relationship between electrides and topological materials. 

\begin{figure}[htp]
	\centering
	\includegraphics[width=0.95\linewidth]{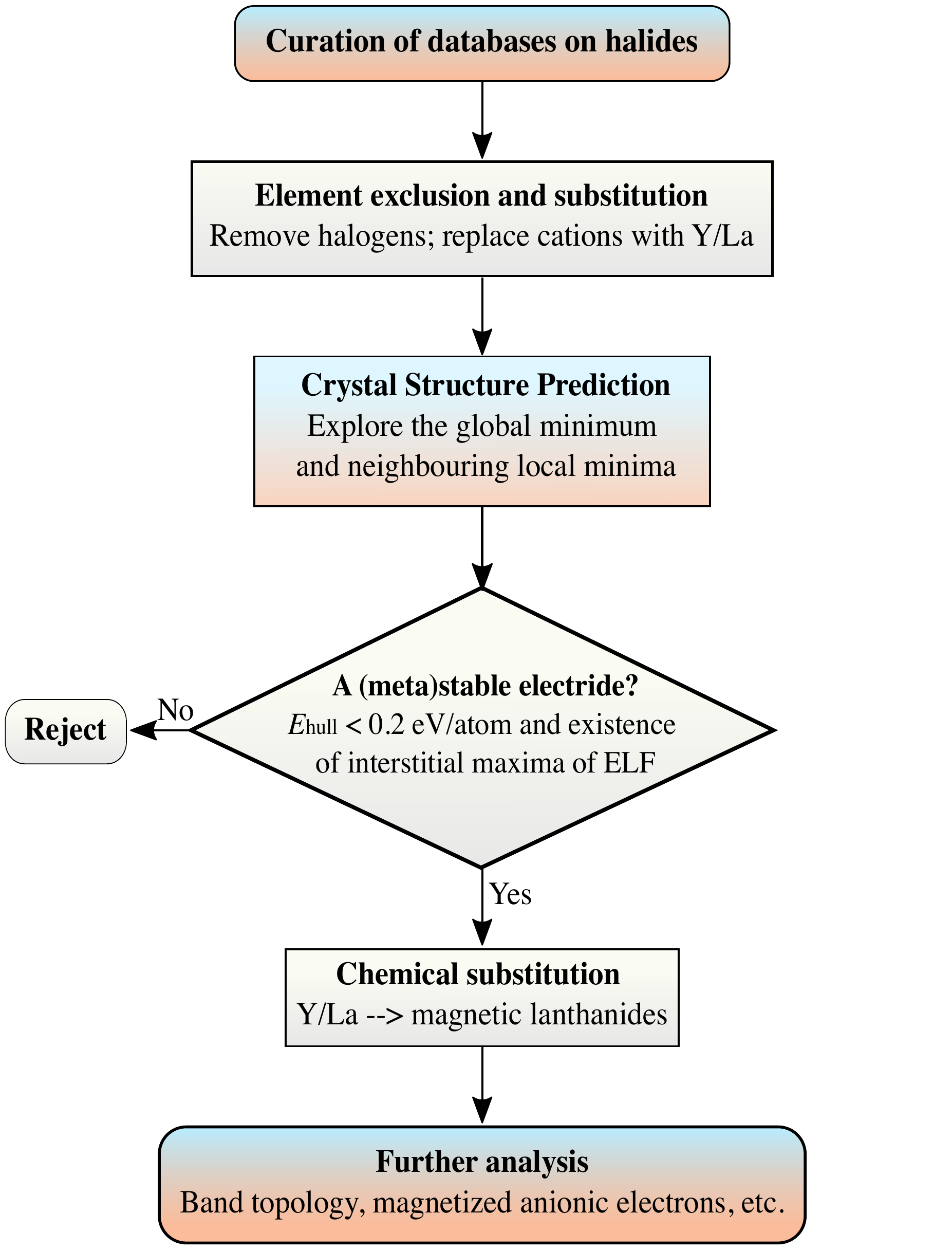}
	\caption{Flowchart of designing halide-derived electrides.}
	\label{flowchart}
\end{figure}

\section{Methods}
\subsection{Design scheme}
\label{subsec_design_scheme}

The procedure of our design scheme is presented in Figure \ref{flowchart}, beginning with the selection of halides as electride hosts. For the sake of concreteness, we pick 11 promising ternary halides as host crystals (see Appendix \ref{sec_allcrystals}), which can be classified into 4 stoichiometric families, $A_2$C$X_p$, $A$C$X_p$, $A_2$Ge$X_p$ and $A$Ge$X_p$, where $A$ and $X$ stand for rare-earth and halogen elements, respectively; $p$ be the ratio of mole fraction of the halogen. The halogen atoms in these materials are surrounded by cations, so that the exclusion of halogens gives rise to an intrinsic excess of electrons as well as interstitial sites to accommodate those excess electrons, both of which are necessary for electride formation \cite{zhu2019computational}. We have restricted our scope to rare-earth based halides, considering that rare-earth elements are favorable electron donors in electrides because of their low eletronegativity and standard valence (usually +3). Moreover, close radii among rare-earth atoms form a good condition for chemical substitution (e.g. Y $\rightarrow$ Gd). Owing to intermediate electronegativities (2.55 and 2.01, respectively), C and Ge from group \rom{4}A are chosen as anionic elements, so that electrons could deviate from cationic atoms, and reside in interstitial orbitals at the same time \cite{wang2017exploration}. Si with the electronegativity 1.90 is not considered here, as it is less electronegative compared to C and Ge. In fact, electrons are barely attracted by rare-earth atoms even in the case of Ge, see Sec. \ref{subsec_overview}. Thus, the consideration of Si, which is less likely to attract electrons, is obsolete. We point out that the materials screening here is merely based on rules of thumb, which cannot rule out the possibility of potential electride hosts that disobey these conditions, but would always be an efficient guide of searching electrides \cite{zhang2017computer}.

Given the selected halides, we generate initial structures for the CSP. Taken away the halogens, the corresponding 4 categories of binary compounds are $A_2$C, $A$C, $A_2$Ge and $A$Ge, or collectively designated by $A_nB$ ($n=1,2; B=$ C, Ge). Further, the original cation element $A$ is replaced with Y and La separately to develop 2 groups of compositions for each category. For clarity and consistency, throughout the paper we will insist on a classification in four stoichiometric categories, i.e. the member of one category is independent of the chemical nature of $A$. Structure optimization and comparison would be conducted on the Y and La groups of materials individually for each category of $A_nB$. Since neither of Y and La has partially filled 4\textit{f} orbitals, the CSP is entirely freed from spin-polarized calculations to reduce the computational costs. Another reason for this replacement is related to the chemical substitution of magnetic lanthanides, which will be elaborated below.

CSP then follows to eliminate the structure instability caused by the exclusion of halogen atoms. From the viewpoint of CSP, the reduced structure of a known material acts as a reasonable initial guess, assisting CSP to reach the ground state more efficiently, in contrast to random starting structures \cite{zhang2017computer}. Notice that although the size of interstitial regions might be compressed during the structure optimization, proper vacant spaces adopting anionic electrons are often found among energy-low-lying structures in practice. 

Subsequent to the CSP, convex hull diagrams are constructed in order to estimate the stability and potential synthesizability of predicted structures relative to competing phases with different stoichiometries. The most stable structures whose energies above the convex hull (denoted as $E_\text{hull}$) are smaller than 0.20 eV/atom \cite{oganov2019structure, wang2017exploration} are processed for electride identification, in order not to rule out the synthesis possibility of metastable polymorphs \cite{PhysRevLett.116.075503, sun1600225thermodynamic}. Under each category of $A_nB$, the first five promising electrides with the lowest energies are kept as final candidates for both $A=$ Y and La. We note that $E_\text{hull}$ of these candidates mostly turn out to be smaller than 0.10 eV/atom (see Sec. \ref{subsec_overview}). 

Next, for every stoichiometric category of potential electrides $A_nB$, the cations of 10 candidate structures (5 from Y-based systems and 5 from the La-based) are substituted by magnetic lanthanides (we focus on Gd in this work) so that magnetism could be induced. As the ionic radius of Gd is between the ones of Y and La, it is expedient to assume stable or metastable Gd-based phases could be found among structures inherited from those of Y and La, especially in the case that Y$_nB$ and La$_nB$ share an identical crystal structure. This is the other reason why Y and La are simultaneously introduced at the stage of generating initial structures. In addition to the close radii of substitution elements, the stability is also assessed by constructing convex hulls. For simplicity, only ferromagnetic (FM) configurations of each Gd-based electride candidates are computed. At the same time, anionic electrons easily become spin-unpaired in the case of FM configurations of Gd. Applied to all other lanthanides with in-between ionic radii, this element substitution strategy could massively explore stable or metastable phases of lanthanide-based compounds without an expensive CSP for every lanthanide. This is a key point of our work.

Finally, having all the predicted electrides, we concentrate on the structures of interest for detailed analysis, including the dynamical stability, further confirmation of electride features, etc. The magnetic ground state of selected Gd systems is computed (see Sec. \ref{subsec_comp_methods}), estimating the relative stability of the FM state compared to other possible magnetic states. Also, we particularly characterize the topological nature encoded in electronic bands of the designed electrides. Notice that these procedures are applied to only a subset of predicted crystals, as given in the last box of Figure \ref{flowchart}.

\subsection{Computational methods} 
\label{subsec_comp_methods}

All the electronic structure calculations are performed with the density functional theory (DFT) based on the projector augmented wave scheme \cite{PhysRevB.59.1758} as implemented in the Vienna \textit{ab initio} simulation package ({\sc vasp}) \cite{PhysRevB.54.11169}. The exchange-correlation functional is given by the generalized gradient approximation (GGA) of the Perdew-Burke-Ernzerhof (PBE) type \cite{PhysRevLett.77.3865}. Strong intra-atomic interactions in 4\textit{f} orbitals of Gd are approximated by the GGA+U \cite{PhysRevB.52.R5467}. The nonlocal Heyd-Scuseria-Ernzerhof (HSE06) hybrid functional is supplemented to check the band crossing \cite{heyd2003hybrid}. The energy cutoff of the plane wave is chosen as $30\%$ above of the maximum value recommended in the pseudopoetential library. The Brillouin zone is sampled by a $\Gamma$-centered grid with the resolution of $2\pi \times 0.03 \text{ \AA}^{-1}$. All atoms are fully relaxed with a force criterion of 0.001~eV/{\AA}. Pre/Post-processing tools and utilities for solids computation \cite{wang2019vaspkit, herath2020107080, setyawan2010299, togo2018texttt, momma2008vesta} are employed.

Crystal structure searches for various compositions of electrides are performed with the minima hopping (MH) \cite{goedecker2004minima} algorithm with a generalized version  to periodic systems employing variable cell shape molecular dynamics \cite{amsler2010crystal,flores2020crystal}. This is a well-tested methodology that aims at finding the global minimum of a complex condensed matter system \cite{flores2020perspective}. For crystals with a small number ($<10$) of atoms per primitive cell, the supercells composed of two even three primitive cells are included as well to extend the searching space. As aforementioned, not only the lowest energy configuration, but also neighboring local minima are evaluated for their potential as electrides. In order to build convex hulls for final candidates, stable phases recorded in Materials Project \cite{Jain2013Materials, Ong2008Li} are adopted. The formation energy is estimated with respect to the corresponding stable elemental polymorph, where the {\sc vasp} pseudopotential C\_$\,$h is employed for all carbon-based systems, as some $A_n$C compounds contain $C_2$ dimers; the plane wave energy cutoffs are 875.0 eV and 320.6 eV for C- and Ge- based materials, respectively. To assess the lattice dynamics, we use the supercell approach, on which the total energy and pattern of displacements (within the harmonic approximation) are evaluated with {\sc vasp}. The harmonic interatomic force constants are built and diagonalized using the {\sc phonopy} software \cite{togo2015first}.

We utilize a recently developed scheme termed CMP+SDFT \cite{huebsch2020benchmark}, that is based on a combination of the cluster multipole (CMP) theory \cite{PhysRevB.95.094406, PhysRevB.99.174407} and calculations in the framework of spin-density functional theory (SDFT), in order to explore the possible stable magnetic states for Gd-based candidates of interest. The CMP basis is an orthogonal basis set of magnetic multipole configurations that can be used to expand an arbitrary magnetic structure. By an analysis of experimental data, it has been shown that the most stable magnetic configurations in nature are linear combinations of only few CMPs. Hence, an exhaustive list of candidate magnetic configurations can be generated with the initial magnetic moments generated by CMPs and their combinations. Subsequently, SDFT calculations are exploited to seek the stable configurations. In the case that the FM state is a local minimum instead of a global one, the energy difference higher than the global minimum is employed to evaluate the energetic stability of the FM structure. We point out again that the magnetic ground state is determined only for a subset of designed magnetic electrides.

\begin{figure*}[htp]
	\centering
	\includegraphics[width=0.98\linewidth]{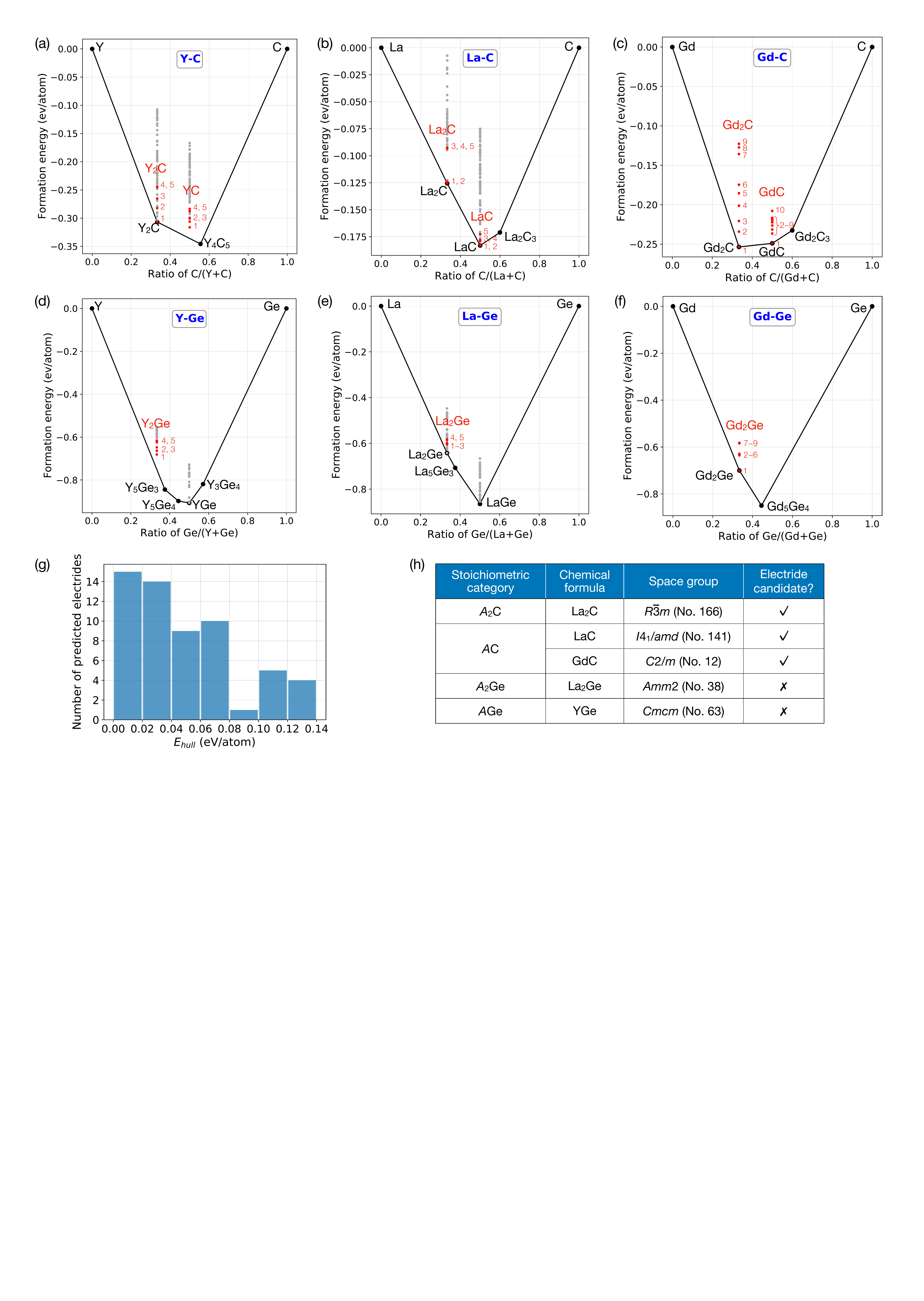}
	\caption{A statistical overview of predicted materials. (a)-(f) Convex hull diagrams of binary compounds $A_nB$ ($A=$ Y, La, Gd; $B=$ C, Ge). Electride candidates are colored by red, and distinguished by the attached numbers that represent their energy rankings. Notice that the number 1 means the corresponding phase has the lowest formation energy merely within the predicted electride structures; there could be a non-electride structure that is more stable. Grey dots stand for the other (meta)stable structures predicted by the CSP, which showing no evidence of electrides or have relatively high formation energy. No grey dots in diagrams of the Gd systems, since the Gd-based structures are obtained by chemical substitution instead of the CSP. (g) Stability (measured by $E_\text{hull}$) distribution of 58 designed electrides. (h) New stable polymorphs ($E_\text{hull} = 0$) of each stoichiometric category discovered by the CSP. Non-electride systems are also presented.}
	\label{overvew}
\end{figure*}

Along the high-symmetry line, crystalline symmetries of energy bands are determined via \texttt{irvsp}, a software package that can obtain irreducible representations of the electronic Bloch states \cite{gao2020irvsp}. Tight-binding models in Wannier representation \cite{wu2018wanniertools} are constructed through Wannier functions without the iterative maximal-localization procedure \cite{PhysRevB.56.12847, PhysRevB.65.035109, pizzi2020wannier90}. Considering the non-nucleus-bound electronic states are mostly \textit{s}-like \cite{matsuishi2003high}, we set the \textit{s} orbital as the trial projection at interstitial sites.  

\subsection{Electrides identification}
The electron localization function (ELF) \cite{becke1990simple, savin2005electron} is one of the mature theoretical descriptors of electrides \cite{liu2020electrides, zhu2019computational, zhao2016electride, dale2018theoretical, wang2017exploration}. The emergence of ELF local maxima (irregardless of its magnitude \cite{zhu2019computational}) at the vacant region indicates a potential electride. For the further identification of candidates of interest, we do the fatband analysis, namely, projecting eigenstates onto the \textit{s} orbital of empty spheres (pseudoatoms) located at the vacancy positions with ELF maxima. The wavefunction on the interstitial-state dominating bands as well as partial charge density close to the Fermi energy represent additional proof.
\vspace{1.5cm}

\section{Results}

\subsection{Overview of predicted electrides}
\label{subsec_overview}

\begin{figure*}[tbp]
	\centering
	\includegraphics[width=0.75\linewidth]{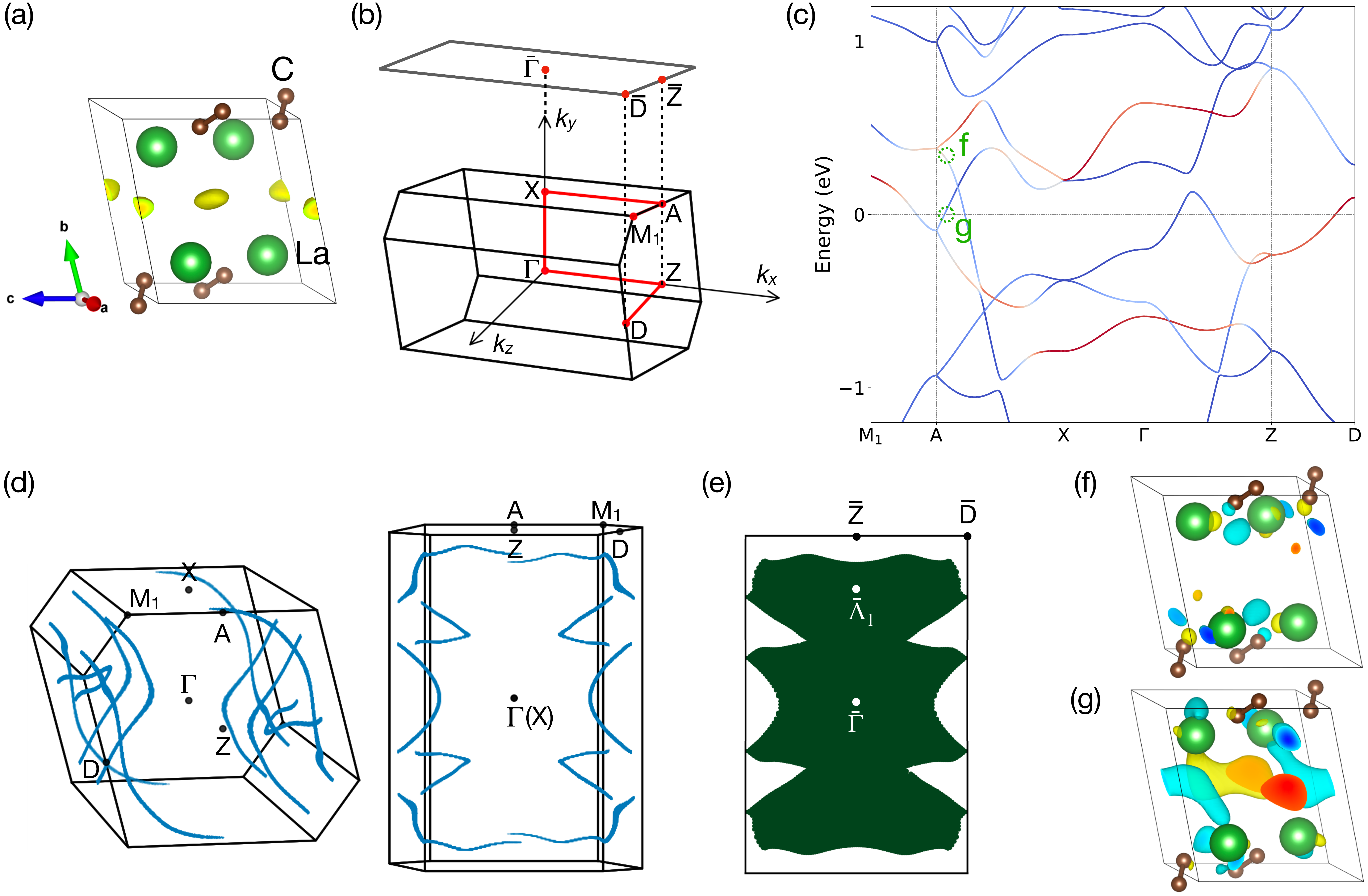}
	\caption{Electronic properties from the bulk calculation of LaC. (a) Primitive cell of LaC. Yellow bubbles represent the ELF. For a clear visualization, the ELF around atoms is not displayed. (b) The Brillouin zone of LaC. The high-symmetry points and the path chosen are highlighted in red. The $y$-direction projected Brillouin zone is indicated too. (c) Projected band structure of LaC along the path denoted in (b). The Fermi energy is set 0. As the color changes from red to blue, the contribution of interlayer states diminishes. (d) Nodal line of LaC in the reciprocal space. The left and right panels are viewed from \textit{x} and \textit{y} directions respectively. (e) The Zak phase integrated along the \textit{y} direction. Green and white separately stand for $\pi$ and 0 values. High-symmetry points in this reduced reciprocal space are also shown. (f)(g) Real branches of wavefunctions with the corresponding positions indicated in (c). The yellow and cyan mean positive and negative parts, respectively; red and blue stand for their cross-sections.}
	\label{LaC_bulk}
\end{figure*}

Within the 4 stoichiometric categories of binary compounds, $A_2$C, $A$C, $A_2$Ge and $A$Ge ($A=$ Y, La, Gd), 58 stable or metastable phases of potential electrides are identified. Among the former three categories, 56 of them are identified for the first time and hence, have not been recorded previously in the Inorganic Crystal Structure Database (ICSD) \cite{bergerhoff1983inorganic}. In $A$Ge no electride appears within the range of $E_\text{hull} < 0.20$ eV/atom, implying that Ge is less favorable for electride phases than C in the rare-earth based systems, which might be due to the smaller electronegativity. Figure \ref{overvew} (a)-(f) illustrate the distribution of predicted stable or metastable structures in convex hull diagrams. The potential electrides are highlighted in red, ranked and labeled according to their formation energies from low to high. The (meta)stable phases that are disregarded because of either showing no evidence to be electrides or yielding a high formation energy compared to other structures are marked in grey. As given in Figure \ref{overvew}(g), all the 58 electride structures have the $E_\text{hull}$ smaller than 0.14 eV/atom, and 49 ($84.5\%$) of them are below 0.10 eV/atom, revealing the possible synthesizability of these candidate structures \cite{oganov2019structure}.

Figure \ref{overvew}(h) lists the the newly discovered materials corresponding to the most stable phase ($E_\text{hull} = 0$). Two non-electride systems are attached too for reference. In the category $A_2$C, being the ground state, La$_2$C with the anti-CdCl$_2$ structure is predicted to be a promising electride. Note that the identical structure is shared by the stable phases of the Y$_2$C and the FM Gd$_2$C, while the latter two systems have been experimentally confirmed as stable 2D electrides \cite{zhang2014two, lee2020ferromagnetic}. Additionally, the FM configuration is verified to be the energetically favorable magnetic state of Gd$_2$C according to CMP+SDFT, in agreement with experiments \cite{lee2020ferromagnetic}. Therefore, the stable phases of $A_2$C represent a benchmark of our design scheme, indicating the repeatability of our CSP strategy. Moreover, the presumption of chemical substitution in Sec. \ref{subsec_design_scheme} is demonstrated to hold, i.e. stable or metastable Gd-based structures could be derived from those of Y and La.

We proceed by introducing the other stable structures in Figure \ref{overvew}(h). In the stoichiometric family $A$C, the possible electrides LaC with the space group $I4_1/amd$ (No. 141) and the FM GdC with the space group $C2/m$ (No. 12) are revealed to be the ground states, neither of which was reported previously. For the category $A_2$Ge, we find an unreported global minimum structure La$_2$Ge that belongs to the space group $Amm2$ (No. 38), though it is unlikely to be an electride. Finally, in the category $A$Ge, despite the absence of electride possibility amongst the most stable phases, a new stable polymorph of YGe is found, sharing the same structure with the known orthorhombic LaGe. Crystal structures of these stable materials are presented in Appendix \ref{sec_stable_structures}.

\begin{figure}[tbp]
	\centering
	\includegraphics[width=0.95\linewidth]{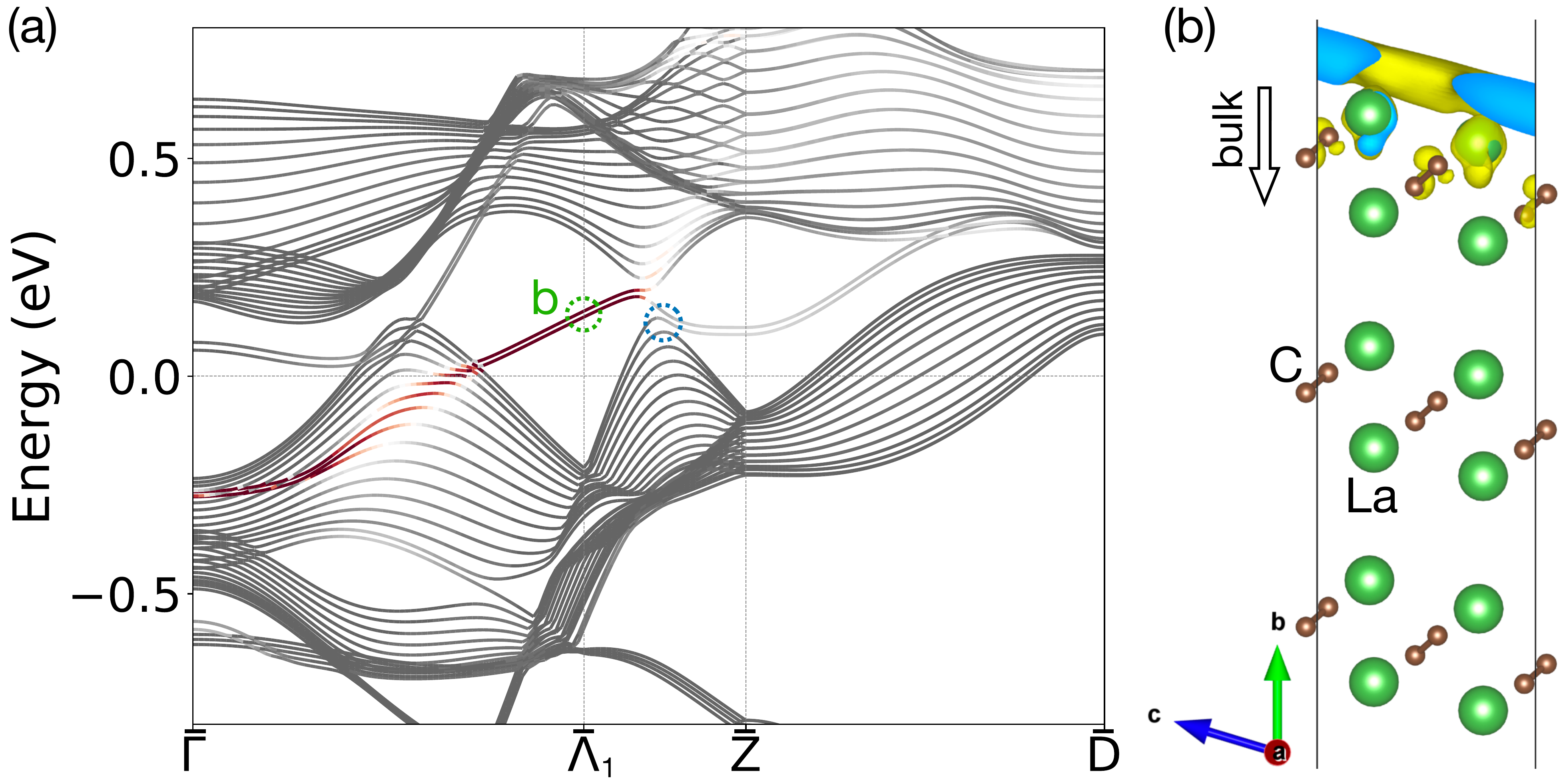}
	\caption{Electronic structure of a 15-layer $(010)$ LaC slab. (a) Electronic band structure of the LaC slab. High-symmetry points in the reduced Brillouin zone are given by Figure \ref{LaC_bulk}(b) or \ref{LaC_bulk}(e). Energy is measured from the Fermi level. The red indicates a large proportion of floating electronic states that localize on the top and bottom of the slab. (b) The squared wavefunction of a surface state localized around the top layer. The corresponding point in the band structure is marked by a green circle in (a).}
	\label{LaC_surf}
\end{figure}

Relative to the corresponding stable phases, metastable phases can exhibit superior physical and chemical characteristics \cite{sun1600225thermodynamic}. In the subsequent discussion, we shall focus on two metastable systems to exemplify the crystal structures and electronic properties of the new electride candidates. One system (denoted as the $A$C system) is represented by the monoclinic LaC with $E_\text{hull} = 4.0$ meV/atom, being a promising 2D electride. The other (denoted as the $A_2$Ge system) is based on an orthorhombic structure, found in Y$_2$Ge and La$_2$Ge concurrently. These two electride systems will be thoroughly discussed in Sec. \ref{subsec_AC} and Sec. \ref{subsec_A2Ge}, respectively.

\subsection{The \textit{A}C system} 
\label{subsec_AC}

\begin{figure*}[htp]
	\centering
	\includegraphics[width=0.7\linewidth]{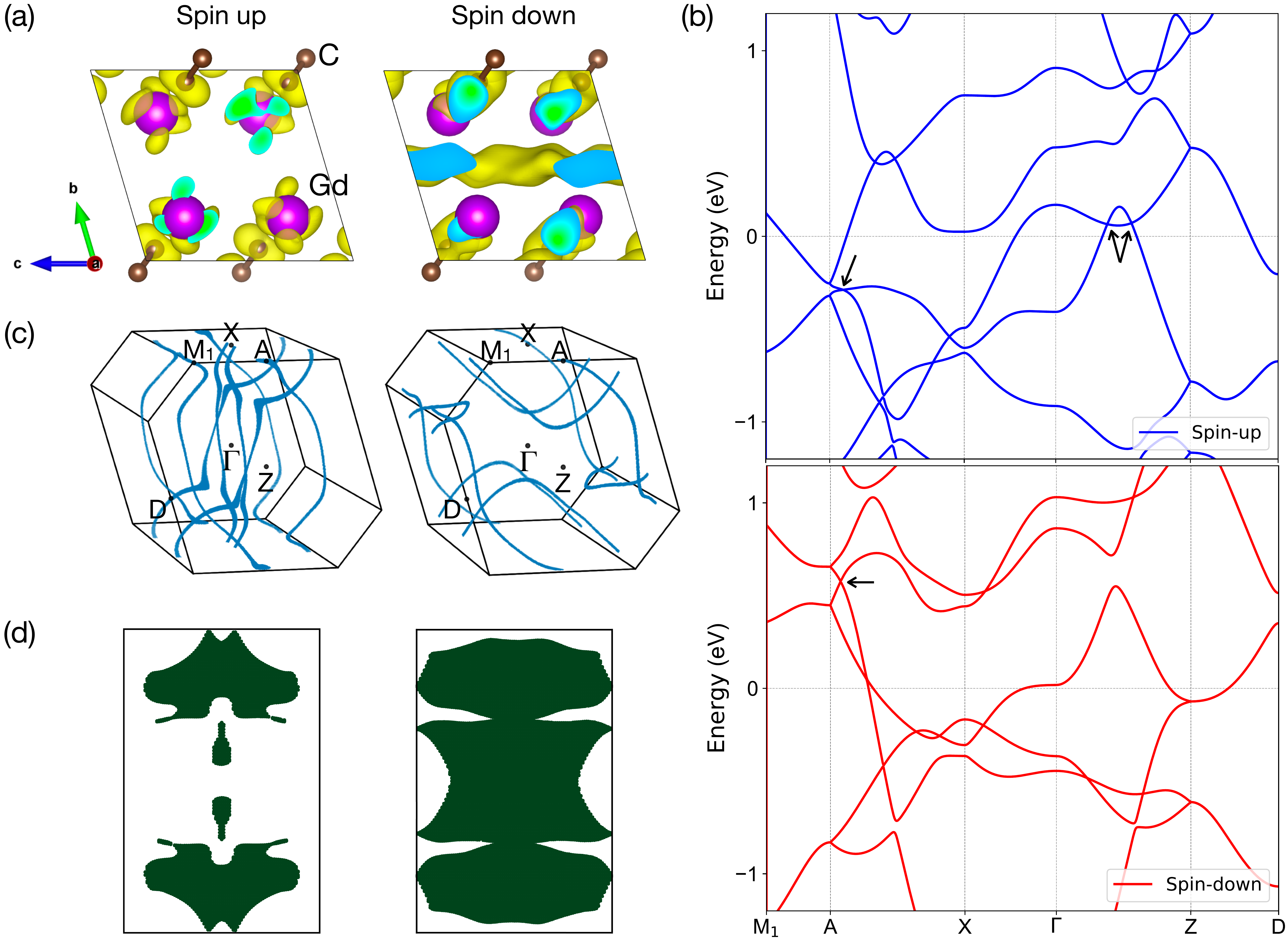}
	\caption{Electronic properties of GdC. (a) Primitive cell of GdC. Partial charge densities around the Fermi level ($-0.05 \sim 0.05$ eV) are shown for spin up (left) and spin down (right). Yellow bubbles represent the partial charge density, the green and blue are the cross sections. (b) Electronic band structure of FM GdC for spin up (top) and spin down (bottom). The high-symmetry points are given in (c). Arrows are indicative of the degeneracies between the highest VB and the lowest CB. (c) Nodal lines of FM GdC for the spin-up (left) and spin-down (right) channels. The Brillouin zone together with high-symmetry points are depicted too. (d) The Zak phase for spin up (left) and spin down (right), integrated along the \textit{y} direction, i.e. $\Gamma$-$X$ direction. Green and white mean $\pi$ and 0 values, respectively.}
	\label{GdC_bulk}
\end{figure*}

Starting from Y$_2$Br$_2$C$_2$ \cite{simon1996supraleitung}, we predict a metastable nonmagnetic crystal LaC , with the formation energy $-0.18$ eV/atom, $4.0$ meV/atom above the convex hull, showing no imaginary phonon modes (see Appendix \ref{sec_phonon}). This is an original material not reported in the database, whose crystal structure is shown in Figure \ref{LaC_bulk}(a). LaC crystallizes in a monoclinic lattice with the space group $P2_1/c$ (No.14), and is composed of four La atoms and four C atoms in a primitive cell. Aside from the time reversal ($\mathcal{T}$) symmetry, the following two symmetries are essential: an inversion $\mathcal{P}$ whose center is at the corner of the unit cell, and an \textit{a}-axis twofold screw operation $\mathcal{S}_{2[100]}$. Indicated by the ELF, LaC is a promising 2D electride, where excess electrons are trapped between cationic layers [Figure \ref{LaC_bulk} (a)]. 

We calculate the electronic band structure along the high-symmetry path given in Figure \ref{LaC_bulk}(b), in the absence of spin-orbit coupling (SOC). Entangled bands around the Fermi level reveal the metallic behavior of LaC. Placing pseudoatoms with the Wigner-Seitz radius 1.50 \AA$\ $at positions with local ELF maxima, i.e. fractional coordinates (0, 0.5, 0) and (0.5, 0.5, 0.5), yields the contribution of anionic electrons to the Bloch wavefunctions. One can notice that interstitial states are closely related to bands around the Fermi energy $E_\text{F}$ and show conducting behavior. The number of valence electrons in this calculation is 60. Hence, the 30 lower bands are counted as valence bands (VBs), while the other higher bands are regarded as conduction bands (CBs). In Figure \ref{LaC_bulk} (f)(g), the real parts of two eigenstates affiliated to the highest VB and the lowest CB are depicted respectively, of which the corresponding positions are marked in the band structure of Figure \ref{LaC_bulk}(c), close to a gapless point. Non-nuclear orbitals are observed for the VB, further confirming the electride characteristics. Nevertheless, this VB claimed by the projection analysis to have little portion of interstitial states, also contains non-nuclear orbitals according to the wavefunction visualization [Figure \ref{LaC_bulk}(g)]. Such contradiction originates from the fact that those non-nuclear orbitals are accidentally distributed in the uncovered region of pseudoatoms. The projection with empty spheres therefore underestimates the contribution from anionic electrons.

Next we move to the topic of band topology. The gapless point between a VB and a CB, mentioned in the last paragraph, is located along the $A$-$X$ line that is invariant under the $\mathcal{S}_{2[100]}$ operation. This degeneracy is protected by the opposite $\mathcal{S}_{2[100]}$ eigenvalues of the two crossing bands, indicated by \texttt{irvsp}. Additionally, the VB and CB remain gapless under the HSE06 hybrid functional. Owing to the $\mathcal{PT}$ symmetry in this spinless system \cite{chan20163, fang2016topological}, this degenerate point extends to general \textit{k}-points and forms a snake-shape nodal line [Figure \ref{LaC_bulk}(d)]. In the manifold of the Brillouin zone, the disconnected nodal line segments should constitute a whole loop. It is noteworthy that this nodal line partly comes from non-nucleus-bound electronic states (e.g. the band crossing in the $A$-$X$ direction), which are insensitive to the relativistic effect of the nuclei field, therefore might exhibit a negligibly small gap even if SOC is present \cite{zhang2019topological}. 

The nodal line here is protected by the $\pi$ Berry phase, defined as
\begin{equation}
	\label{Zak_phase}
	\gamma(\boldsymbol{k}_{\parallel}) = \sum_{n}^{\text{occ.}} \int_{0}^{|\boldsymbol{K}|}\mathrm{d}k_{\perp}\ i\langle u_{n\boldsymbol{k}} | \nabla_{k_{\perp}} | u_{n\boldsymbol{k}} \rangle
\end{equation}
for an arbitrary wave vector $\boldsymbol{k}$, where $\boldsymbol{K}$ stands for the reciprocal lattice vector, $k_{\perp}$ is the magnitude of $\boldsymbol{k}$ component along $\boldsymbol{K}$ and $\boldsymbol{k}_{\parallel}$ is the other component perpendicular to $\boldsymbol{K}$; $|u_{n\boldsymbol{k}}\rangle$ represents the cell-periodic Bloch function, with the gauge $|u_{n,\boldsymbol{k}+\boldsymbol{K}}\rangle = e^{-i\boldsymbol{K}\cdot\boldsymbol{r}}|u_{n\boldsymbol{k}}\rangle$. Note that the Berry phase specified in Eq. (\ref{Zak_phase}), with the integral path along a reciprocal lattice vector, is also known as the Zak phase \cite{PhysRevLett.62.2747}. Making $\boldsymbol{K}$ parallel with the \textit{y} direction of the reciprocal space, yields the Zak phase distribution displayed in Figure \ref{LaC_bulk}(e), which is quantized to be 0 or $\pi$ (mod $2\pi$) as a result of the $\mathcal{P}\mathcal{T}$ symmetry. The nontriviality of this nodal line is demonstrated by the fact that the profile of $\pi$ value region of the Zak phase just matches the nodal line projection along the \textit{y} direction [Figure \ref{LaC_bulk}(e) and the right panel of Figure \ref{LaC_bulk}(d)]. LaC is therefore a TNLE \cite{hirayama2018electrides}.

While underlying the fingerprint of nodal lines in the bulk, the Zak phase also discloses information on the surface. To elaborate this point, we construct a 15-layer (010) slab with the electron layer terminated, and compute the electronic band structure using DFT, as presented in Figure \ref{LaC_surf}(a), whose high-symmetry points in the surface Brillouin zone are given by Figure \ref{LaC_bulk}(b) or \ref{LaC_bulk}(e). Contribution from excess electrons localized on the surfaces is highlighted by projecting eigenstates onto pseudoatoms on the top and bottom ends of the slab. At the wave vector $\boldsymbol{k}_{\parallel}$ with $\pi$ Zak phase [e.g., $\bar{\Lambda}_1$ in Figure \ref{LaC_bulk}(b)], the bulk polarization is nonzero and quantized, and the topological charge $e/2$ (mod $e$) would be induced at the slab surface by this bulk polarization \cite{PhysRevB.48.4442, hirayama2017topological}. As the two surfaces of the slab are equivalent on account of the $\mathcal{P}$ symmetry, the total charge at $\bar{\Lambda}_1$ is different from that in the region with 0 Zak phase [e.g., $\bar{Z}$ in Figure \ref{LaC_bulk}(b)] by \textit{e}. The charge difference can be evidenced by the slab band structure. Along the $\bar{Z}\rightarrow\bar{\Gamma}$ direction, one unoccupied state traverses the bulk line node and merges into the valence spectrum in the end, as marked by the blue circle in Figure \ref{LaC_surf}(a). As a result, in comparison with $\bar{Z}$, there exists one more VB at $\bar{\Lambda}_1$, being responsible for the charge difference \textit{e}. In addition to the surface charge, midgap states emerging from the line node are observed. On the left side of the line node, the midgap states are dominated by interstitial orbitals. In Figure \ref{LaC_surf}(b), visualization of a midgap state in real space unveils the fact that this boundary state is not tightly attached to any atoms, but floats on the surface instead. Such floating surface states are also predicted in other TNLEs like Y$_2$C \cite{hirayama2018electrides}.

\begin{figure*}[htbp]
	\centering
	\includegraphics[width=0.75\linewidth]{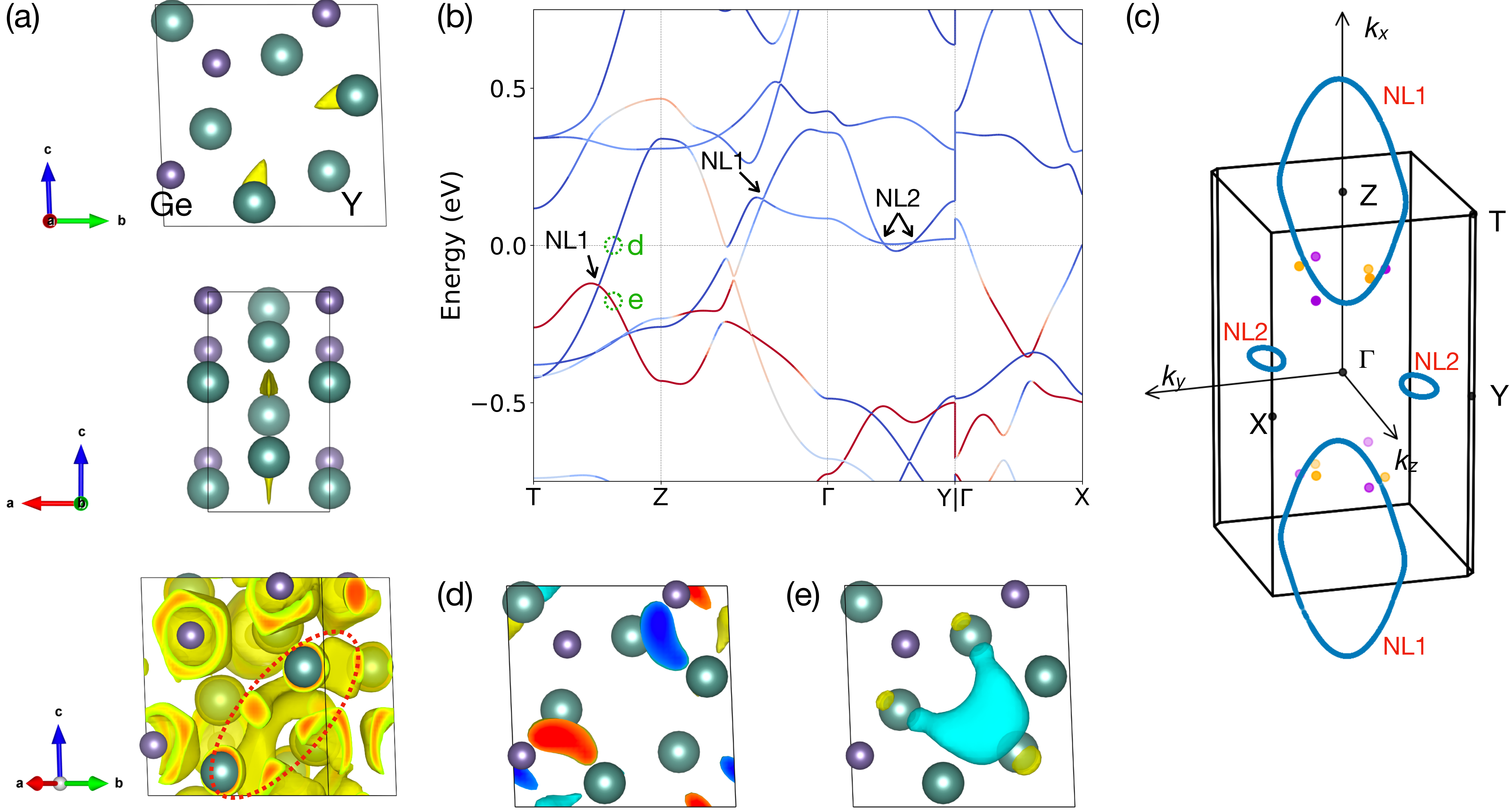}
	\caption{Electronic properties from the bulk calculation of Y$_2$Ge. (a) Primitive cell of Y$_2$Ge. Yellow bubbles are ELF. The isosurface value in the bottom panel is smaller than those in the top and middle panels. Connected interstitial states are highlighted by a dashed red circle in the bottom panel. (b) Projected band structure of Y$_2$Ge with the high-symmetry points denoted in (c). The Fermi energy is set 0. As the color changes from red to blue, the contribution of interstitial states decays. Some gapless points are indicated by arrows. (c) Nodal lines of Y$_2$Ge accompanied by the Brillouin zone. Two inequivalent groups of nodal lines are labels as NL1 and NL2, respectively. Spinless Weyl points are also displayed, with the orange and magenta representing the monopole charge $+1$ and $-1$ separately. (d)(e) show the real branches of the wavefunctions at the positions indicated in (b); the positive part is in yellow and the corresponding cross-sections are colored by red; the cyan and blue are for the negative part.}
	\label{Y2Ge_bulk}
\end{figure*}

We now discuss the crystal after chemical substitution Y $\rightarrow$ Gd, namely, the isostructural GdC. As described in Sec. \ref{subsec_design_scheme}, we have restricted our discussion to the FM configuration of Gd atoms, in which case the interstitial electrons can be easily magnetized. For the GdC here, although an antiferromagnetic (AFM) configuration is predicted by CMP+SDFT to be the magnetic ground state, it is not highly energetic-favorable compared to the FM state (see Appendix \ref{sec_ground_mag}). This makes our 'FM restriction' more plausible. GdC possesses a formation energy equal to $-0.22$ eV/atom and $E_\text{hull}= 0.032$ eV/atom. To look into the electronic structure, the Hubbard $U$ for Gd 4\textit{f} orbitals is included, with $U - J = 6$ eV and $J = 0.7$ eV \cite{PhysRevX.4.031023}; other values ($U - J = 3$ eV and 9 eV) are also checked, showing that only the dense flat $4f$ bands shift while other bands near the Fermi level remain unchanged. Figure \ref{GdC_bulk}(a) presents the partial charge density of the FM GdC, in the vicinity of $E_{\text{F}}$ ($-0.05 \sim 0.05$ eV) for two spin channels. While only intralayer electrons can be seen for spin up, excess electrons rest in the interlayer gap for spin down. Therefore, spin-polarized interstitial states are achieved in this FM system, although the macroscopic magnetism is dominated by Gd atoms. Figure \ref{GdC_bulk}(b) gives the electronic band structure of the FM GdC without SOC. One can notice that bands of two spin channels resemble those in LaC, as a result of the close electronic configuration between Y and Gd, and the nearly identical crystal structure. Such a similarity has been also observed in the electride systems Y$_2$C and Gd$_2$C \cite{PhysRevX.4.031023, liu2020ferromagnetic}. Hence, we expect GdC exhibits a similar electronic behavior to LaC, especially regarding the band topology. Band degeneracies around the Fermi level are displayed in Figure \ref{GdC_bulk}(b), all of which are preserved at the level of HSE06 functionals. In parallel with LaC, these degenerate points turn out to extend in the reciprocal space, forming nodal rings for both spin up and down [Figure \ref{GdC_bulk}(c)]. All the degeneracies still exist in the HSE06 band structure. The nontriviality is again verified by the Zak phase [Figure \ref{GdC_bulk}(d)]. In spite of the broken $\mathcal{T}$ symmetry, the remaining $\mathcal{P}$ symmetry still protects the $\pi$ Berry phase nodal lines. This time reversal breaking TNLE state was also reported in Gd$_2$C \cite{liu2020ferromagnetic}.

\subsection{The \textit{A}$_2$Ge system}
\label{subsec_A2Ge}

\begin{figure*}[htbp]
	\centering
	\includegraphics[width=0.7\linewidth]{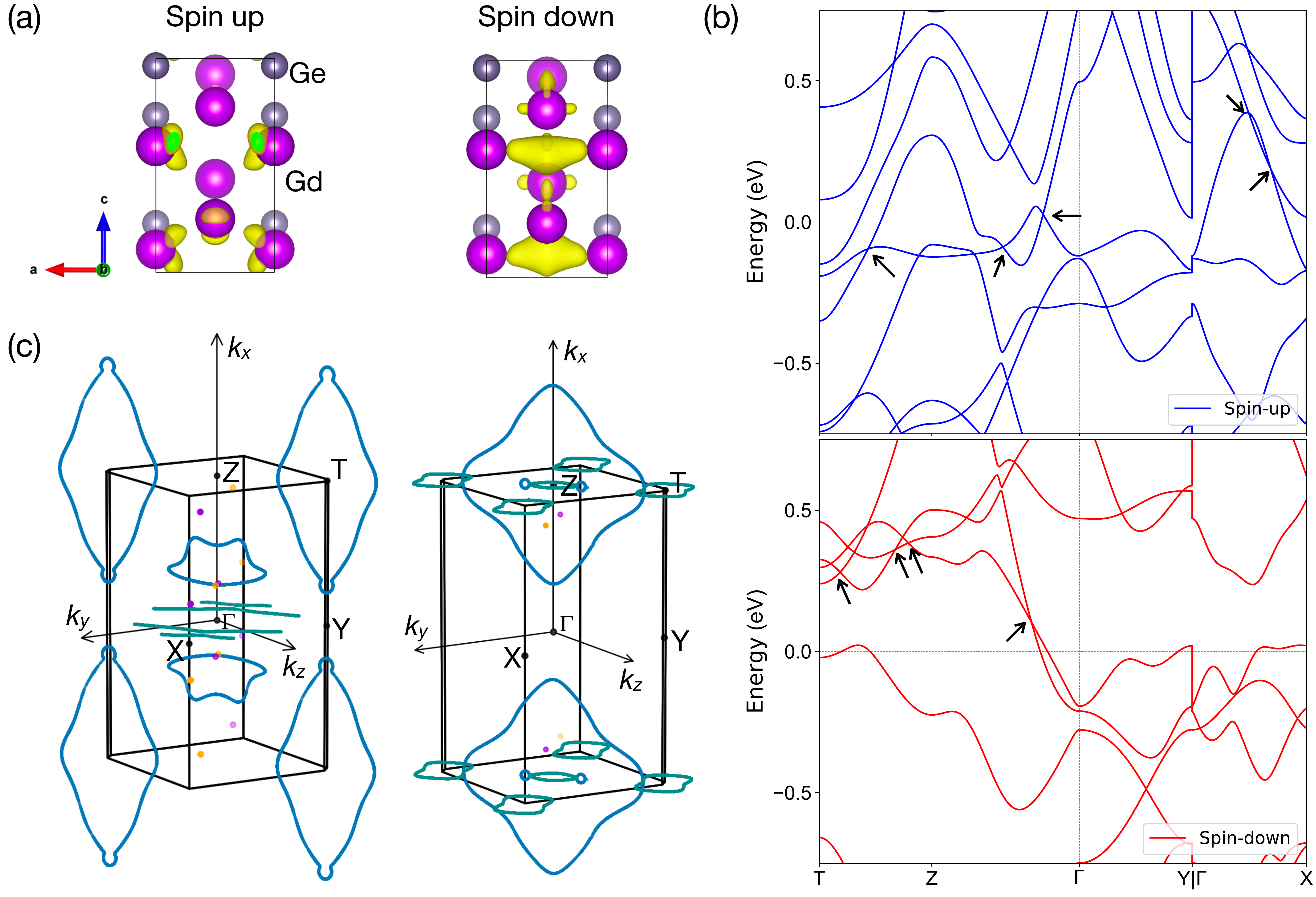}
	\caption{Electronic properties of Gd$_2$Ge. (a) Primitive cell of Gd$_2$Ge. Partial charge densities around the Fermi level ($-0.05 \sim 0.05$ eV) are presented for spin up (left) and spin down (right). The yellow bubbles represent the partial charge density, and the green means the cross-section. (b) Electronic band structure of FM GdC for the spin-up (top) and spin-down (bottom) parts. The high-symmetry points are indicated in (c). Degenerate points between the highest VB and lowest CB are marked by arrows. (c) Nodal lines of FM GdC for spin up (left) and spin down (right). The Brillouin zone, together with high-symmetry points, are depicted too. The blue or green distinguishes nodal lines that are distributed on the vertical or horizontal mirror plane. Weyl points are also displayed, with the orange and magenta representing the monopole and antimonopole, respectively.}
	\label{Gd2Ge_bulk}
\end{figure*}

Y$_2$Ge is another new inorganic electride, derived from the halide La$_2$I$_2$Ge\cite{mattausch2005reduced}, with the structure metastability supported by the negative formation energy $-0.62$ eV/atom ($0.13$ eV/atom higher than the hull) and the absence of imaginary phonon frequencies (see Appendix \ref{sec_phonon}). The base-centered orthorhombic lattice of Y$_2$Ge belongs to the space group $Amm2$ (No. 38). The same structure emerges as a metastable phase of La$_2$Ge too. Figure \ref{Y2Ge_bulk}(a) depicts the primitive cell of Y$_2$Ge that contains nine Y atoms and three Ge atoms. The symmetry operations are: a twofold rotation $\mathcal{C}_{2[0\bar{1}1]}$, two mirror reflections $\mathcal{M}_{[011]}$ and $\mathcal{M}_{[100]}$, whose axes are perpendicular to each other. Notice that the inversion symmetry $\mathcal{P}$ is absent. The spatial distribution of the ELF is given in Figure \ref{Y2Ge_bulk}(a). Two ELF maxima related by the $\mathcal{C}_{2[0\bar{1}1]}$ or  $\mathcal{M}_{[011]}$ symmetry exist in the cationic cavities (the top and middle panels). These two interstitial sites are not isolated but connected with each other, which is uncovered in the bottom panel with a smaller isosurface value of ELF. Now that these interstitial states are connected through the one-dimensional (1D) channel, we classify this candidate as a 1D electride. Set pseudoatoms with the Wigner-Seitz radius 0.5 \AA$\ $at the interstitial locations with local maxima of the ELF, yielding the projected band structure of Y$_2$Ge (without SOC) given by Figure \ref{Y2Ge_bulk}(b), in which the high-symmetry points are denoted in Figure \ref{Y2Ge_bulk}(c). Eigenstates around the Fermi level mostly emanate from Y $4d$ and interstitial \textit{s} orbitals. Again the number of VBs is selected to match the half number of valence electrons in the primitive cell.

Along the $T$-$Z$-$\Gamma$-$Y$ path, the gap between the highest VB and the lowest CB disappears at four positions, which have been labeled by arrows NL1 or NL2 in Figure \ref{Y2Ge_bulk}(b). While the crossings marked by NL1 on the $T$-$Z$ and $Z$-$\Gamma$ paths are confirmed by the band structure computed with the HSE06 hybrid functional, the other two crossings NL2 on the $\Gamma$-$Y$ path disappear when employing HSE06 (see Appendix \ref{sec_hse_bands}). For completeness, we take the latter two degenerate points into account too. Since the $T$-$Z$-$\Gamma$-$Y$ line is located on the $\mathcal{M}_{[011]}$-invariant plane $(011)$, each state on this line has a well-defined eigenvalue $\pm1$ of $\mathcal{M}_{[011]}$, as verified by \texttt{irvsp}. It turns out that every pair of crossing bands have opposite mirror eigenvalues, protecting the degeneracy. In Figure \ref{Y2Ge_bulk}(d)(e), two eigenstates---one is reflection antisymmetric and the other is symmetric with respect to the (011) plane---are depicted. Their corresponding positions have been indicated in Figure \ref{Y2Ge_bulk}(b), belonging to a CB and a VB, respectively. Such opposite crystalline symmetries guarantee the band crossing in the $T$-$Z$ direction. Moreover, one of the crossing bands is dominated by anionic electrons [Figure \ref{Y2Ge_bulk}(e)], again suggesting the significant role of interstitial states for band inversion in electrides. Figure \ref{Y2Ge_bulk}(c) reveals the nodal lines stemming from those gapless points are confined within the mirror plane $(011)$. The big ring (labeled as NL1) crosses the boundary of the Brillouin zone; two small rings (labeled as NL2) are the inversion of each other, associated by the $\mathcal{T}$ symmetry. They are related to the band crossings NL1 and NL2 in Figure \ref{Y2Ge_bulk}(b), respectively. As the degenaracies on the $\Gamma$-$Y$ line disappear at the HSE06 level, the small nodal rings NL2 might be absent under an experimental observation. We then implement the $\mathbb{Z}$-type reflection invariant to characterize the mirror-symmetry protected nodal lines \cite{PhysRevB.90.205136, fang2016topological}. Within the mirror plane, we pick two points $p_a$ and $p_b$ on the two sides of the nodal line, at which the VBs and CBs are separated. Then count the number of VBs with the mirror eigenvalue $+1$, denoted by $N_{a,b}$. The reflection invariant reads
\begin{equation}
	\label{Z-invariant}
	\zeta = N_a - N_b.
\end{equation}
A nonzero $\zeta$ confirms the nontrivial nodal line, which means that the band inversion cannot be adiabatically eliminated without breaking the mirror symmetry. In our case, we obtain $\zeta_1 = \zeta_2 = 1$ for the big and small nodal rings. As a result, the stability of the line nodes in Y$_2$Ge is protected by the reflection symmetry $\mathcal{M}_{[011]}$.

Aside from the line nodes, there appear six pairs of point nodes (spinless Weyl points) owing to the absence of $\mathcal{P}$ symmetry. We compute the monopole charge (also known as the chirality) of each node using the Berry flux across an enclosed surface accommodating the node \cite{doi:10.7566/JPSJ.87.041002}. Figure \ref{Y2Ge_bulk}(c) depicts the positions and monopole charges of the Weyl points. Each pair of Weyl points are linked by the $\mathcal{T}$ symmetry, possessing the same monopole charge.

Next, we replace the Y with Gd while retaining the crystal structure, obtaining a metastable magnetic electride whose formation energy is $-0.58$ eV/atom and $E_\text{hull}=0.12$ eV/atom. Figure \ref{Gd2Ge_bulk}(a) displays the partial charge density distribution in the primitive cell, within the energy range $-0.5\text{ eV} < E-E_{\text{F}} < 0.5\text{ eV}$. Similar to GdC, interstitial states prefer the minority spin (spin down) in Gd$_2$Ge. That is, the magnetization opposite to Gd is induced among anionic electrons. Utilizing the GGA+U method, we show the electronic band structure in Figure \ref{Gd2Ge_bulk}(b) without SOC. Compared to the bands of Y$_2$Ge, the spin-up bands are lowered and the spin-down bands are lifted as expected. Degenerate points along high-symmetry lines are labeled by arrows. All the degenaracies are confirmed by the computation using the HSE06 functional, except for the right two band crossings on the $T$-$Z$ path of the spin-down band structure, which are replaced by one crossing (see Appendix \ref{sec_hse_bands}). These gapless points are guaranteed by the mirror symmetries $\mathcal{M}_{[011]}$ and $\mathcal{M}_{[100]}$. Figure \ref{Gd2Ge_bulk}(c) presents the nodal lines generated from the band crossings within the mirror planes. In contrast to Y$_2$Ge, the $k_x=0$ (spin up)  and $k_x=\pi$ (spin down) planes also hold nodal lines (highlighted in green), allowed by the $\mathcal{M}_{[100]}$ symmetry. Additionally, the perpendicular mirror planes give rise to two sets of nodal chains in the reciprocal space of spin down. In parallel with Y$_2$Ge, Weyl points are observed for both spin channels [Figure \ref{Gd2Ge_bulk}(c)]. The intriguing properties arose out of the coexistence of Weyl points and nodal chains remain to be explored.

\section{Conclusion} 
In summary, we put forward a new electride design scheme, predicting electrides as the reduced form of halides. To demonstrate the success of this design strategy, three stoichiometric categories of binary compounds, $A_2$C, $A$C, $A_2$Ge ($A$ = Y, La, Gd), with 56 newly discovered stable or metastable phases derived from ternary halides, are identified as promising electrides based on first-principles crystal structure prediction and chemical substitution. Of all the candidate structures, the energies above the convex hull are smaller than $0.14$ eV/atom, indicating potential synthesizability. Benefiting from the similarities of atomic and chemical properties of rare-earth elements, stable or metastable Gd-based systems are generated by element substitution from the Y- and La-based structures so that magnetism can manifest itself. Subsequently, we study the ferromagnetic case, a situation where the anionic electrons can easily be magnetized. Two electride systems, the monoclinic $A$C and the orthorhombic $A_2$Ge, have been discussed in detail. According to our results for $A=$ Gd, their magnetic states are suitable for spin injection. The topological phases investigated show that both of the two designed electride systems manifest topological nodal lines in the reciprocal space, which cements the close relation between topological materials and electrides because of the interstitial states around the Fermi level. 

Our strategy to design magnetic and topological electrides can be easily generalized to other sets of materials.  For example, one could expand the searching space to include hydrides, oxides, nitrides, etc. accessible through material databases in combination with high-throughput techniques \cite{setyawan2010high-throughput}. The replacement of other lanthanides in host systems is likely to open new avenues to synthesize magnetic electrides. One future task would be collaborating with experimentalists to synthesize the designed electrides, and devising incorporation into spintronic devices.

\begin{acknowledgments}
M. H. acknowledges financial support from JSPS KAKENHI Grants No. 20K14390. J. A. F.-L. acknowledges computational resources provided by the Swiss National Supercomputing Center withing the project s970. T. N. acknowledges financial support from JST PRESTO (Grant No. JPMJPR20L7). M.-T. H. and R. A. were supported by a Grant-in-Aid for Scientific Research (No. 19H05825, and No. 16H06345) from the Ministry of Education, Culture, Sports, Science and Technology.	
\end{acknowledgments}

\newpage
\appendix

\section{Initial halides} 
\label{sec_allcrystals}
\begin{figure}[htbp]
	\centering
	\includegraphics[width=\linewidth]{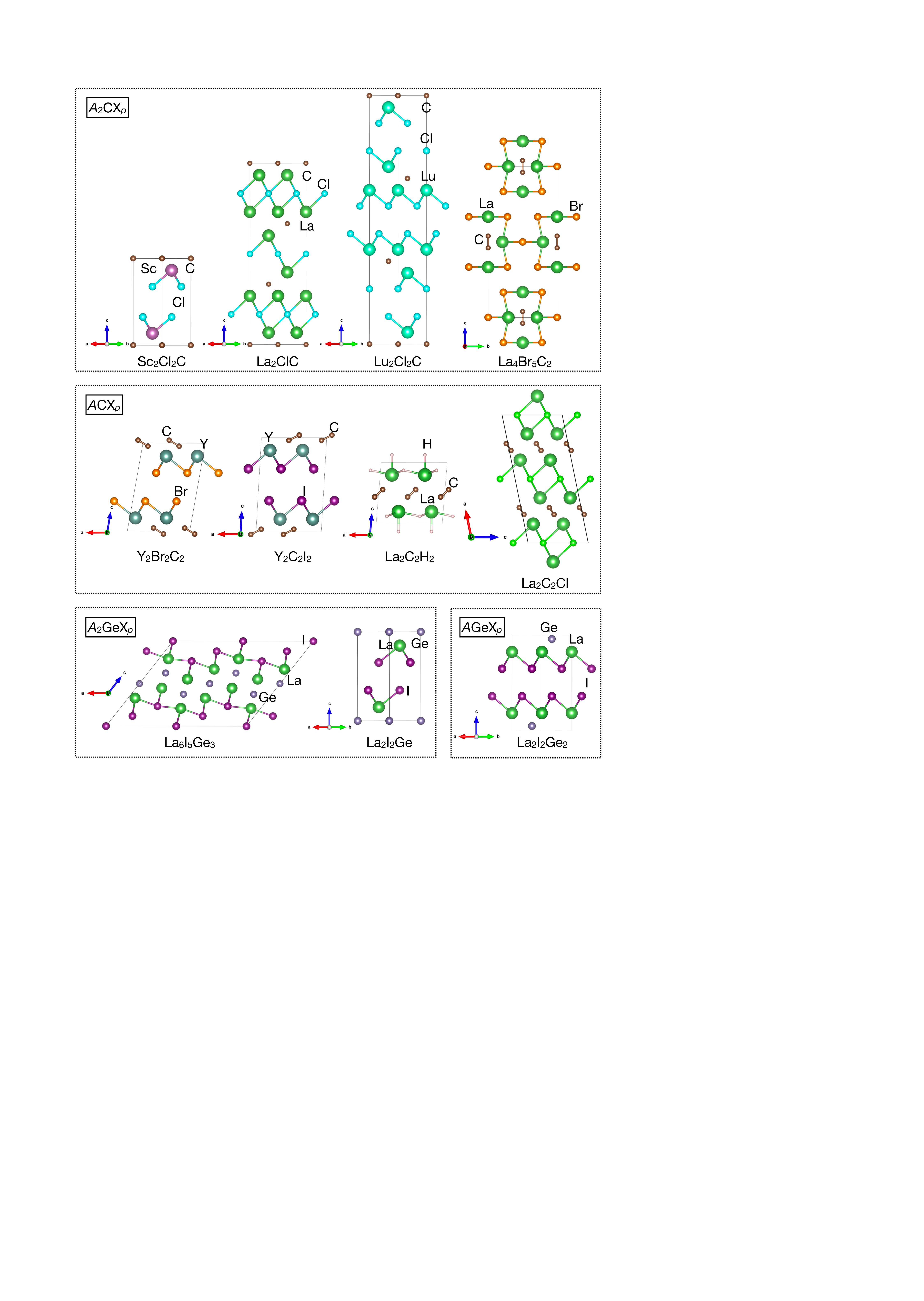}
	\caption{Primitive cells of the initial compounds that serve as electride hosts. Boxes classify the stoichiometric categories. All the compounds are subject to halides, with one exception La$_2$C$_2$H$_2$ (a hydride) that makes no difference after taking away the hydrogens.}
	\label{initial_crystals}
\end{figure}

In this Appendix we list the structures of initial ternary halides as presented in Figure \ref{initial_crystals}. For the crystal whose number of atoms in a primitive cell is smaller than 10 after excluding halogens, like Y$_2$Br$_2$C$_2$ and La$_2$I$_2$Ge$_2$, the doubled even tripled supercell is involved into the CSP in addition to the primitive cell.

\section{Newly discovered stable materials}
\label{sec_stable_structures}
\begin{figure}[htbp]
	\centering
	\includegraphics[width=\linewidth]{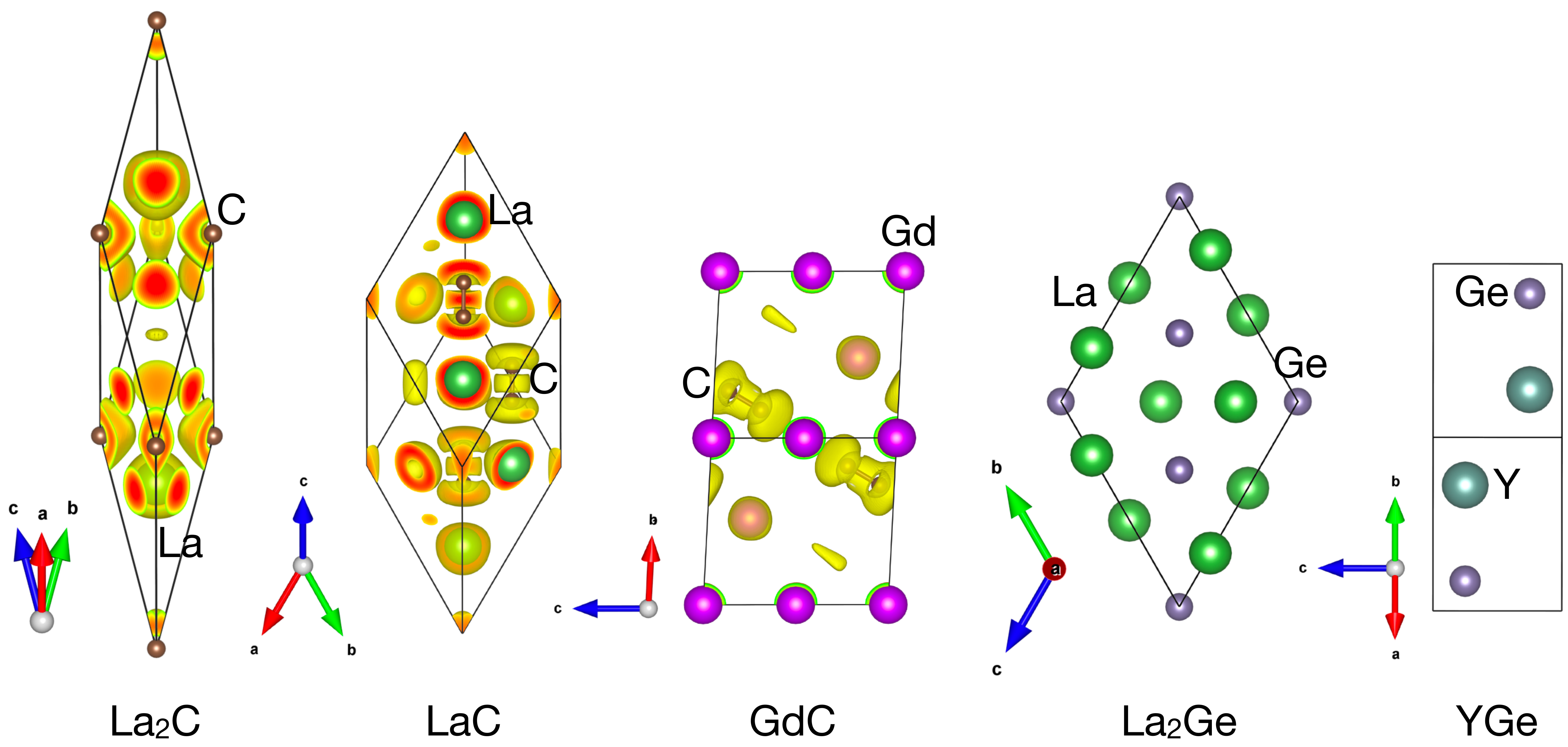}
	\caption{Primitive cells of newly discovered crystals that correspond to the ground polymorphs in accordance with our calculations. The left three are shown to be potential electrides, with their ELF distributions displayed.}
	\label{stable_structures}
\end{figure}

 Figure \ref{stable_structures} illustrates crystals structures of stable phases unveiled by our CSP (see Sec. \ref{subsec_overview}), which are hitherto unexpected under their own chemical compositions.

\section{Phonon density of states\\of LaC and Y$_2$Ge}
\label{sec_phonon}
\begin{figure}[htbp]
	\centering
	\includegraphics[width=\linewidth]{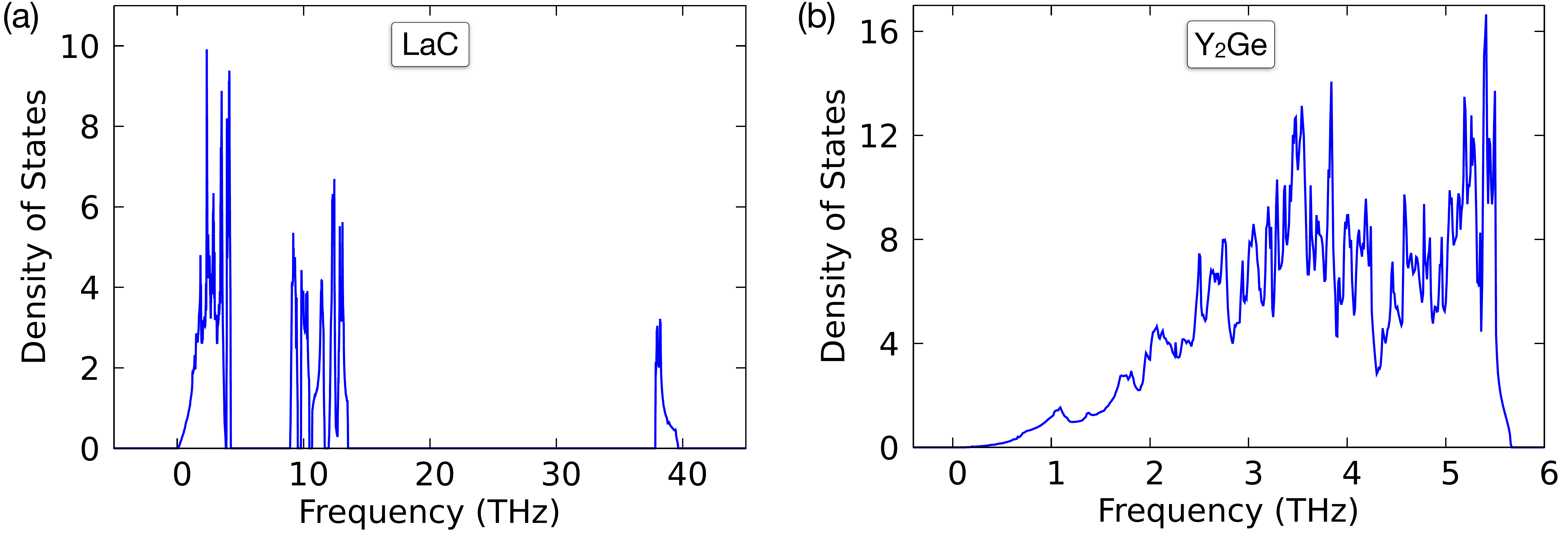}
	\caption{Phonon DOS of (a) LaC and (b) Y$_2$Ge, respectively.}
	\label{phonon_spectra}
\end{figure}

The phonon DOS of LaC and Y$_2$Ge are shown in Figure \ref{phonon_spectra}. Since the imaginary frequency is absent, both systems are dynamically stable, supporting further evidence towards their possible synthesis. 

\section{Ground magnetic states\\of GdC and Gd$_2$Ge} 
\label{sec_ground_mag}
\begin{figure}[htp]
	\centering
	\includegraphics[width=0.8\linewidth]{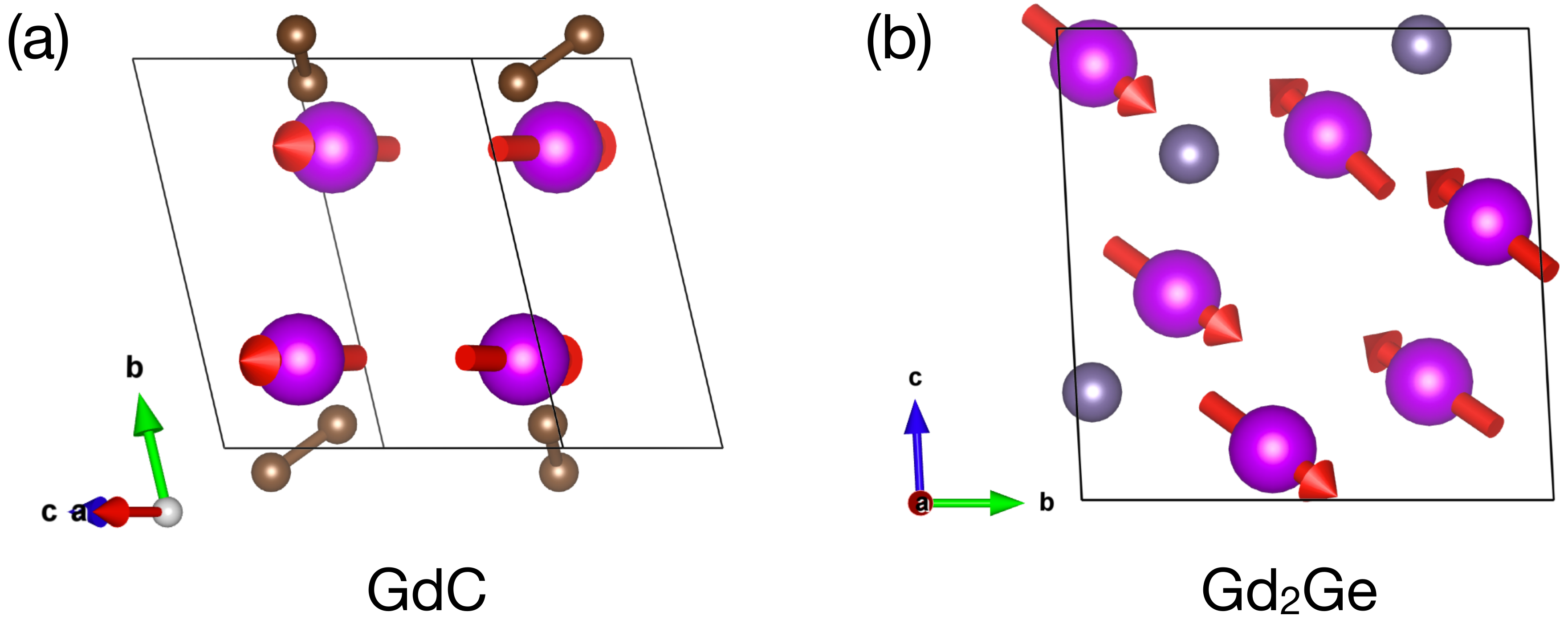}
	\caption{Magnetic ground states of (a) GdC and (b) Gd$_2$Ge, respectively. The red arrow stands for the direction of each spin.}
	\label{ground_mag}
\end{figure}

The ground magnetic configuration of Gd atoms is AFM for both GdC and Gd$_2$Ge structures that are discussed in Sec. \ref{subsec_AC} and \ref{subsec_A2Ge}, as given in Figure \ref{ground_mag}. The total energy of AFM GdC is $0.04$ meV/atom lower than that of the FM case, and energy difference is $4.59$ meV/atom for the Gd$_2$Ge system. 

\section{Band structure with the HSE06 hybrid functional}
\label{sec_hse_bands}
\begin{figure}[htbp]
	\centering
	\includegraphics[width=\linewidth]{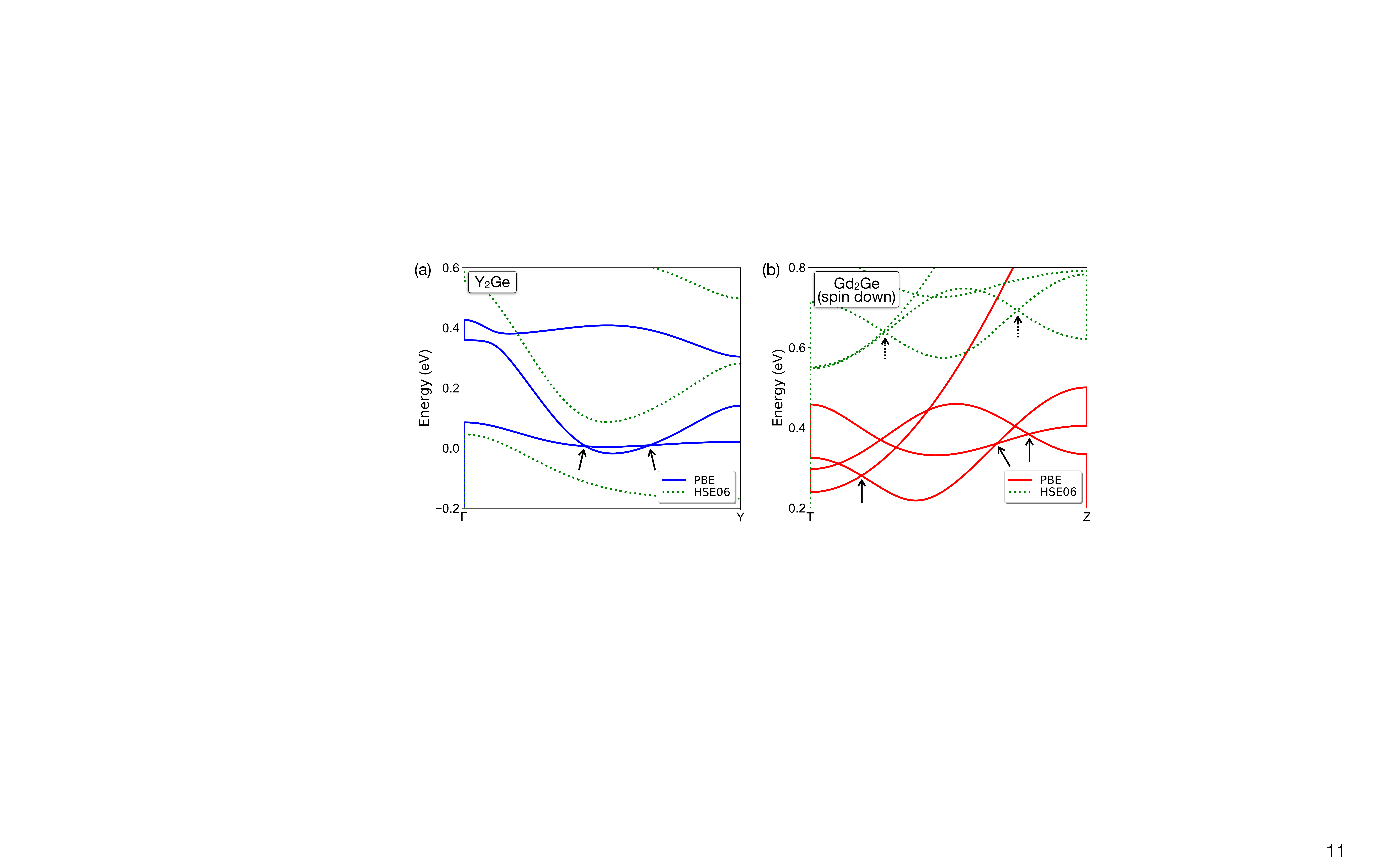}
	\caption{Electronic band structures computed using the PBE pure functional and the HSE06 hybrid functional, indicated by solid and dashed lines, respectively. Degenerate points relating to nodal lines are marked by arrows. (a) is for Y$_2$Ge and (b) is for spin down of Gd$_2$Ge.}
	\label{hse_bands}
\end{figure}
Figure \ref{hse_bands} shows the comparison between electronic band structures at the PBE level and at the HSE06 level. For Y$_2$Ge, two bands along the $\Gamma$-$Y$ path, which are predicted to cross each other by the PBE functional, are shown to be gapped under the nonlocal HSE06 hybrid functional [Figure \ref{Y2Ge_bulk}(b) and Figure \ref{hse_bands}(a)]. In the FM Gd$_2$Ge system, the band structure on the $T$-$Z$ line for the spin-down is depicted in Figure \ref{hse_bands}(b). The right two degenerate points revealed by the PBE functional are substituted by one point at the HSE06 level.

\bibliographystyle{apsrev4-1}
\bibliography{main.bib}

\providecommand{\noopsort}[1]{}\providecommand{\singleletter}[1]{#1}%
\begin{thebibliography}{67}%
\makeatletter
\providecommand \@ifxundefined [1]{%
 \@ifx{#1\undefined}
}%
\providecommand \@ifnum [1]{%
 \ifnum #1\expandafter \@firstoftwo
 \else \expandafter \@secondoftwo
 \fi
}%
\providecommand \@ifx [1]{%
 \ifx #1\expandafter \@firstoftwo
 \else \expandafter \@secondoftwo
 \fi
}%
\providecommand \natexlab [1]{#1}%
\providecommand \enquote  [1]{``#1''}%
\providecommand \bibnamefont  [1]{#1}%
\providecommand \bibfnamefont [1]{#1}%
\providecommand \citenamefont [1]{#1}%
\providecommand \href@noop [0]{\@secondoftwo}%
\providecommand \href [0]{\begingroup \@sanitize@url \@href}%
\providecommand \@href[1]{\@@startlink{#1}\@@href}%
\providecommand \@@href[1]{\endgroup#1\@@endlink}%
\providecommand \@sanitize@url [0]{\catcode `\\12\catcode `\$12\catcode
  `\&12\catcode `\#12\catcode `\^12\catcode `\_12\catcode `\%12\relax}%
\providecommand \@@startlink[1]{}%
\providecommand \@@endlink[0]{}%
\providecommand \url  [0]{\begingroup\@sanitize@url \@url }%
\providecommand \@url [1]{\endgroup\@href {#1}{\urlprefix }}%
\providecommand \urlprefix  [0]{URL }%
\providecommand \Eprint [0]{\href }%
\providecommand \doibase [0]{http://dx.doi.org/}%
\providecommand \selectlanguage [0]{\@gobble}%
\providecommand \bibinfo  [0]{\@secondoftwo}%
\providecommand \bibfield  [0]{\@secondoftwo}%
\providecommand \translation [1]{[#1]}%
\providecommand \BibitemOpen [0]{}%
\providecommand \bibitemStop [0]{}%
\providecommand \bibitemNoStop [0]{.\EOS\space}%
\providecommand \EOS [0]{\spacefactor3000\relax}%
\providecommand \BibitemShut  [1]{\csname bibitem#1\endcsname}%
\let\auto@bib@innerbib\@empty
\bibitem [{\citenamefont {Dye}(1990)}]{dye1990663}%
  \BibitemOpen
  \bibfield  {author} {\bibinfo {author} {\bibfnamefont {J.~L.}\ \bibnamefont
  {Dye}},\ }\href {\doibase 10.1126/science.247.4943.663} {\bibfield  {journal}
  {\bibinfo  {journal} {Science}\ }\textbf {\bibinfo {volume} {247}},\ \bibinfo
  {pages} {663} (\bibinfo {year} {1990})}\BibitemShut {NoStop}%
\bibitem [{\citenamefont {Dye}(2003)}]{dye2003electrons}%
  \BibitemOpen
  \bibfield  {author} {\bibinfo {author} {\bibfnamefont {J.~L.}\ \bibnamefont
  {Dye}},\ }\href {\doibase 10.1126/science.1088103} {\bibfield  {journal}
  {\bibinfo  {journal} {Science}\ }\textbf {\bibinfo {volume} {301}},\ \bibinfo
  {pages} {607} (\bibinfo {year} {2003})}\BibitemShut {NoStop}%
\bibitem [{\citenamefont {Matsuishi}\ \emph {et~al.}(2003)\citenamefont
  {Matsuishi}, \citenamefont {Toda}, \citenamefont {Miyakawa}, \citenamefont
  {Hayashi}, \citenamefont {Kamiya}, \citenamefont {Hirano}, \citenamefont
  {Tanaka},\ and\ \citenamefont {Hosono}}]{matsuishi2003high}%
  \BibitemOpen
  \bibfield  {author} {\bibinfo {author} {\bibfnamefont {S.}~\bibnamefont
  {Matsuishi}}, \bibinfo {author} {\bibfnamefont {Y.}~\bibnamefont {Toda}},
  \bibinfo {author} {\bibfnamefont {M.}~\bibnamefont {Miyakawa}}, \bibinfo
  {author} {\bibfnamefont {K.}~\bibnamefont {Hayashi}}, \bibinfo {author}
  {\bibfnamefont {T.}~\bibnamefont {Kamiya}}, \bibinfo {author} {\bibfnamefont
  {M.}~\bibnamefont {Hirano}}, \bibinfo {author} {\bibfnamefont
  {I.}~\bibnamefont {Tanaka}}, \ and\ \bibinfo {author} {\bibfnamefont
  {H.}~\bibnamefont {Hosono}},\ }\href {\doibase 10.1126/science.1083842}
  {\bibfield  {journal} {\bibinfo  {journal} {Science}\ }\textbf {\bibinfo
  {volume} {301}},\ \bibinfo {pages} {626} (\bibinfo {year}
  {2003})}\BibitemShut {NoStop}%
\bibitem [{\citenamefont {Sui}\ \emph {et~al.}(2019)\citenamefont {Sui},
  \citenamefont {Wang},\ and\ \citenamefont {Duan}}]{sui2019prediction}%
  \BibitemOpen
  \bibfield  {author} {\bibinfo {author} {\bibfnamefont {X.}~\bibnamefont
  {Sui}}, \bibinfo {author} {\bibfnamefont {J.}~\bibnamefont {Wang}}, \ and\
  \bibinfo {author} {\bibfnamefont {W.}~\bibnamefont {Duan}},\ }\href {\doibase
  10.1021/acs.jpcc.8b10379} {\bibfield  {journal} {\bibinfo  {journal} {The
  Journal of Physical Chemistry C}\ }\textbf {\bibinfo {volume} {123}},\
  \bibinfo {pages} {5003} (\bibinfo {year} {2019})}\BibitemShut {NoStop}%
\bibitem [{\citenamefont {Wolf}\ \emph {et~al.}(2001)\citenamefont {Wolf},
  \citenamefont {Awschalom}, \citenamefont {Buhrman}, \citenamefont {Daughton},
  \citenamefont {von Moln{\'a}r}, \citenamefont {Roukes}, \citenamefont
  {Chtchelkanova},\ and\ \citenamefont {Treger}}]{wolf2001spintronics}%
  \BibitemOpen
  \bibfield  {author} {\bibinfo {author} {\bibfnamefont {S.~A.}\ \bibnamefont
  {Wolf}}, \bibinfo {author} {\bibfnamefont {D.~D.}\ \bibnamefont {Awschalom}},
  \bibinfo {author} {\bibfnamefont {R.~A.}\ \bibnamefont {Buhrman}}, \bibinfo
  {author} {\bibfnamefont {J.~M.}\ \bibnamefont {Daughton}}, \bibinfo {author}
  {\bibfnamefont {S.}~\bibnamefont {von Moln{\'a}r}}, \bibinfo {author}
  {\bibfnamefont {M.~L.}\ \bibnamefont {Roukes}}, \bibinfo {author}
  {\bibfnamefont {A.~Y.}\ \bibnamefont {Chtchelkanova}}, \ and\ \bibinfo
  {author} {\bibfnamefont {D.~M.}\ \bibnamefont {Treger}},\ }\href {\doibase
  10.1126/science.1065389} {\bibfield  {journal} {\bibinfo  {journal}
  {Science}\ }\textbf {\bibinfo {volume} {294}},\ \bibinfo {pages} {1488}
  (\bibinfo {year} {2001})}\BibitemShut {NoStop}%
\bibitem [{\citenamefont {Hirayama}\ \emph
  {et~al.}(2018{\natexlab{a}})\citenamefont {Hirayama}, \citenamefont
  {Matsuishi}, \citenamefont {Hosono},\ and\ \citenamefont
  {Murakami}}]{hirayama2018electrides}%
  \BibitemOpen
  \bibfield  {author} {\bibinfo {author} {\bibfnamefont {M.}~\bibnamefont
  {Hirayama}}, \bibinfo {author} {\bibfnamefont {S.}~\bibnamefont {Matsuishi}},
  \bibinfo {author} {\bibfnamefont {H.}~\bibnamefont {Hosono}}, \ and\ \bibinfo
  {author} {\bibfnamefont {S.}~\bibnamefont {Murakami}},\ }\href {\doibase
  10.1103/PhysRevX.8.031067} {\bibfield  {journal} {\bibinfo  {journal} {Phys.
  Rev. X}\ }\textbf {\bibinfo {volume} {8}},\ \bibinfo {pages} {031067}
  (\bibinfo {year} {2018}{\natexlab{a}})}\BibitemShut {NoStop}%
\bibitem [{\citenamefont {Zhu}\ \emph {et~al.}(2019{\natexlab{a}})\citenamefont
  {Zhu}, \citenamefont {Frolov},\ and\ \citenamefont
  {Choudhary}}]{zhu2019computational}%
  \BibitemOpen
  \bibfield  {author} {\bibinfo {author} {\bibfnamefont {Q.}~\bibnamefont
  {Zhu}}, \bibinfo {author} {\bibfnamefont {T.}~\bibnamefont {Frolov}}, \ and\
  \bibinfo {author} {\bibfnamefont {K.}~\bibnamefont {Choudhary}},\ }\href
  {\doibase https://doi.org/10.1016/j.matt.2019.06.017} {\bibfield  {journal}
  {\bibinfo  {journal} {Matter}\ }\textbf {\bibinfo {volume} {1}},\ \bibinfo
  {pages} {1293 } (\bibinfo {year} {2019}{\natexlab{a}})}\BibitemShut {NoStop}%
\bibitem [{\citenamefont {Park}\ \emph {et~al.}(2018)\citenamefont {Park},
  \citenamefont {Kim},\ and\ \citenamefont {Yoon}}]{PhysRevLett.120.026401}%
  \BibitemOpen
  \bibfield  {author} {\bibinfo {author} {\bibfnamefont {C.}~\bibnamefont
  {Park}}, \bibinfo {author} {\bibfnamefont {S.~W.}\ \bibnamefont {Kim}}, \
  and\ \bibinfo {author} {\bibfnamefont {M.}~\bibnamefont {Yoon}},\ }\href
  {\doibase 10.1103/PhysRevLett.120.026401} {\bibfield  {journal} {\bibinfo
  {journal} {Phys. Rev. Lett.}\ }\textbf {\bibinfo {volume} {120}},\ \bibinfo
  {pages} {026401} (\bibinfo {year} {2018})}\BibitemShut {NoStop}%
\bibitem [{\citenamefont {Huang}\ \emph {et~al.}(2018)\citenamefont {Huang},
  \citenamefont {Jin}, \citenamefont {Zhang},\ and\ \citenamefont
  {Liu}}]{huang2018topological}%
  \BibitemOpen
  \bibfield  {author} {\bibinfo {author} {\bibfnamefont {H.}~\bibnamefont
  {Huang}}, \bibinfo {author} {\bibfnamefont {K.-H.}\ \bibnamefont {Jin}},
  \bibinfo {author} {\bibfnamefont {S.}~\bibnamefont {Zhang}}, \ and\ \bibinfo
  {author} {\bibfnamefont {F.}~\bibnamefont {Liu}},\ }\href {\doibase
  10.1021/acs.nanolett.7b05386} {\bibfield  {journal} {\bibinfo  {journal}
  {Nano Letters}\ }\textbf {\bibinfo {volume} {18}},\ \bibinfo {pages} {1972}
  (\bibinfo {year} {2018})}\BibitemShut {NoStop}%
\bibitem [{\citenamefont {Wang}\ \emph
  {et~al.}(2019{\natexlab{a}})\citenamefont {Wang}, \citenamefont {Sui},
  \citenamefont {Gao}, \citenamefont {Duan}, \citenamefont {Liu},\ and\
  \citenamefont {Huang}}]{PhysRevLett.123.206402}%
  \BibitemOpen
  \bibfield  {author} {\bibinfo {author} {\bibfnamefont {J.}~\bibnamefont
  {Wang}}, \bibinfo {author} {\bibfnamefont {X.}~\bibnamefont {Sui}}, \bibinfo
  {author} {\bibfnamefont {S.}~\bibnamefont {Gao}}, \bibinfo {author}
  {\bibfnamefont {W.}~\bibnamefont {Duan}}, \bibinfo {author} {\bibfnamefont
  {F.}~\bibnamefont {Liu}}, \ and\ \bibinfo {author} {\bibfnamefont
  {B.}~\bibnamefont {Huang}},\ }\href {\doibase 10.1103/PhysRevLett.123.206402}
  {\bibfield  {journal} {\bibinfo  {journal} {Phys. Rev. Lett.}\ }\textbf
  {\bibinfo {volume} {123}},\ \bibinfo {pages} {206402} (\bibinfo {year}
  {2019}{\natexlab{a}})}\BibitemShut {NoStop}%
\bibitem [{\citenamefont {Zhang}\ \emph {et~al.}(2019)\citenamefont {Zhang},
  \citenamefont {Fu}, \citenamefont {Jin}, \citenamefont {Dai}, \citenamefont
  {Liu},\ and\ \citenamefont {Yao}}]{zhang2019topological}%
  \BibitemOpen
  \bibfield  {author} {\bibinfo {author} {\bibfnamefont {X.}~\bibnamefont
  {Zhang}}, \bibinfo {author} {\bibfnamefont {B.}~\bibnamefont {Fu}}, \bibinfo
  {author} {\bibfnamefont {L.}~\bibnamefont {Jin}}, \bibinfo {author}
  {\bibfnamefont {X.}~\bibnamefont {Dai}}, \bibinfo {author} {\bibfnamefont
  {G.}~\bibnamefont {Liu}}, \ and\ \bibinfo {author} {\bibfnamefont
  {Y.}~\bibnamefont {Yao}},\ }\href {\doibase 10.1021/acs.jpcc.9b08446}
  {\bibfield  {journal} {\bibinfo  {journal} {The Journal of Physical Chemistry
  C}\ }\textbf {\bibinfo {volume} {123}},\ \bibinfo {pages} {25871} (\bibinfo
  {year} {2019})}\BibitemShut {NoStop}%
\bibitem [{\citenamefont {Nie}\ \emph {et~al.}(2020)\citenamefont {Nie},
  \citenamefont {Qian}, \citenamefont {Gao}, \citenamefont {Fang},
  \citenamefont {Weng},\ and\ \citenamefont {Wang}}]{nie2020application}%
  \BibitemOpen
  \bibfield  {author} {\bibinfo {author} {\bibfnamefont {S.}~\bibnamefont
  {Nie}}, \bibinfo {author} {\bibfnamefont {Y.}~\bibnamefont {Qian}}, \bibinfo
  {author} {\bibfnamefont {J.}~\bibnamefont {Gao}}, \bibinfo {author}
  {\bibfnamefont {Z.}~\bibnamefont {Fang}}, \bibinfo {author} {\bibfnamefont
  {H.}~\bibnamefont {Weng}}, \ and\ \bibinfo {author} {\bibfnamefont
  {Z.}~\bibnamefont {Wang}},\ }\href {https://arxiv.org/abs/2012.02203}
  {\bibfield  {journal} {\bibinfo  {journal} {arXiv preprint arXiv:2012.02203}\
  } (\bibinfo {year} {2020})}\BibitemShut {NoStop}%
\bibitem [{\citenamefont {Liu}\ \emph {et~al.}(2020{\natexlab{a}})\citenamefont
  {Liu}, \citenamefont {Nikolaev}, \citenamefont {Ren},\ and\ \citenamefont
  {Burton}}]{liu2020electrides}%
  \BibitemOpen
  \bibfield  {author} {\bibinfo {author} {\bibfnamefont {C.}~\bibnamefont
  {Liu}}, \bibinfo {author} {\bibfnamefont {S.~A.}\ \bibnamefont {Nikolaev}},
  \bibinfo {author} {\bibfnamefont {W.}~\bibnamefont {Ren}}, \ and\ \bibinfo
  {author} {\bibfnamefont {L.~A.}\ \bibnamefont {Burton}},\ }\href {\doibase
  10.1039/D0TC01165G} {\bibfield  {journal} {\bibinfo  {journal} {J. Mater.
  Chem. C}\ }\textbf {\bibinfo {volume} {8}},\ \bibinfo {pages} {10551}
  (\bibinfo {year} {2020}{\natexlab{a}})}\BibitemShut {NoStop}%
\bibitem [{\citenamefont {Inoshita}\ \emph {et~al.}(2014)\citenamefont
  {Inoshita}, \citenamefont {Jeong}, \citenamefont {Hamada},\ and\
  \citenamefont {Hosono}}]{PhysRevX.4.031023}%
  \BibitemOpen
  \bibfield  {author} {\bibinfo {author} {\bibfnamefont {T.}~\bibnamefont
  {Inoshita}}, \bibinfo {author} {\bibfnamefont {S.}~\bibnamefont {Jeong}},
  \bibinfo {author} {\bibfnamefont {N.}~\bibnamefont {Hamada}}, \ and\ \bibinfo
  {author} {\bibfnamefont {H.}~\bibnamefont {Hosono}},\ }\href {\doibase
  10.1103/PhysRevX.4.031023} {\bibfield  {journal} {\bibinfo  {journal} {Phys.
  Rev. X}\ }\textbf {\bibinfo {volume} {4}},\ \bibinfo {pages} {031023}
  (\bibinfo {year} {2014})}\BibitemShut {NoStop}%
\bibitem [{\citenamefont {Tada}\ \emph {et~al.}(2014)\citenamefont {Tada},
  \citenamefont {Takemoto}, \citenamefont {Matsuishi},\ and\ \citenamefont
  {Hosono}}]{Tada2014High-Throughput}%
  \BibitemOpen
  \bibfield  {author} {\bibinfo {author} {\bibfnamefont {T.}~\bibnamefont
  {Tada}}, \bibinfo {author} {\bibfnamefont {S.}~\bibnamefont {Takemoto}},
  \bibinfo {author} {\bibfnamefont {S.}~\bibnamefont {Matsuishi}}, \ and\
  \bibinfo {author} {\bibfnamefont {H.}~\bibnamefont {Hosono}},\ }\href
  {\doibase 10.1021/ic501362b} {\bibfield  {journal} {\bibinfo  {journal}
  {Inorganic Chemistry}\ }\textbf {\bibinfo {volume} {53}},\ \bibinfo {pages}
  {10347} (\bibinfo {year} {2014})}\BibitemShut {NoStop}%
\bibitem [{\citenamefont {Burton}\ \emph {et~al.}(2018)\citenamefont {Burton},
  \citenamefont {Ricci}, \citenamefont {Chen}, \citenamefont {Rignanese},\ and\
  \citenamefont {Hautier}}]{Burton2018High-Throughput}%
  \BibitemOpen
  \bibfield  {author} {\bibinfo {author} {\bibfnamefont {L.~A.}\ \bibnamefont
  {Burton}}, \bibinfo {author} {\bibfnamefont {F.}~\bibnamefont {Ricci}},
  \bibinfo {author} {\bibfnamefont {W.}~\bibnamefont {Chen}}, \bibinfo {author}
  {\bibfnamefont {G.-M.}\ \bibnamefont {Rignanese}}, \ and\ \bibinfo {author}
  {\bibfnamefont {G.}~\bibnamefont {Hautier}},\ }\href {\doibase
  10.1021/acs.chemmater.8b02526} {\bibfield  {journal} {\bibinfo  {journal}
  {Chemistry of Materials}\ }\textbf {\bibinfo {volume} {30}},\ \bibinfo
  {pages} {7521} (\bibinfo {year} {2018})}\BibitemShut {NoStop}%
\bibitem [{\citenamefont {Tsuji}\ \emph {et~al.}(2016)\citenamefont {Tsuji},
  \citenamefont {Dasari}, \citenamefont {Elatresh}, \citenamefont {Hoffmann},\
  and\ \citenamefont {Ashcroft}}]{tsuji2016structural}%
  \BibitemOpen
  \bibfield  {author} {\bibinfo {author} {\bibfnamefont {Y.}~\bibnamefont
  {Tsuji}}, \bibinfo {author} {\bibfnamefont {P.~L. V.~K.}\ \bibnamefont
  {Dasari}}, \bibinfo {author} {\bibfnamefont {S.~F.}\ \bibnamefont
  {Elatresh}}, \bibinfo {author} {\bibfnamefont {R.}~\bibnamefont {Hoffmann}},
  \ and\ \bibinfo {author} {\bibfnamefont {N.~W.}\ \bibnamefont {Ashcroft}},\
  }\href {\doibase 10.1021/jacs.6b09067} {\bibfield  {journal} {\bibinfo
  {journal} {Journal of the American Chemical Society}\ }\textbf {\bibinfo
  {volume} {138}},\ \bibinfo {pages} {14108} (\bibinfo {year}
  {2016})}\BibitemShut {NoStop}%
\bibitem [{\citenamefont {Ming}\ \emph {et~al.}(2016)\citenamefont {Ming},
  \citenamefont {Yoon}, \citenamefont {Du}, \citenamefont {Lee},\ and\
  \citenamefont {Kim}}]{ming2016first}%
  \BibitemOpen
  \bibfield  {author} {\bibinfo {author} {\bibfnamefont {W.}~\bibnamefont
  {Ming}}, \bibinfo {author} {\bibfnamefont {M.}~\bibnamefont {Yoon}}, \bibinfo
  {author} {\bibfnamefont {M.-H.}\ \bibnamefont {Du}}, \bibinfo {author}
  {\bibfnamefont {K.}~\bibnamefont {Lee}}, \ and\ \bibinfo {author}
  {\bibfnamefont {S.~W.}\ \bibnamefont {Kim}},\ }\href {\doibase
  10.1021/jacs.6b05586} {\bibfield  {journal} {\bibinfo  {journal} {Journal of
  the American Chemical Society}\ }\textbf {\bibinfo {volume} {138}},\ \bibinfo
  {pages} {15336} (\bibinfo {year} {2016})}\BibitemShut {NoStop}%
\bibitem [{\citenamefont {Zhang}\ \emph {et~al.}(2017)\citenamefont {Zhang},
  \citenamefont {Wang}, \citenamefont {Wang}, \citenamefont {Zhang},\ and\
  \citenamefont {Ma}}]{zhang2017computer}%
  \BibitemOpen
  \bibfield  {author} {\bibinfo {author} {\bibfnamefont {Y.}~\bibnamefont
  {Zhang}}, \bibinfo {author} {\bibfnamefont {H.}~\bibnamefont {Wang}},
  \bibinfo {author} {\bibfnamefont {Y.}~\bibnamefont {Wang}}, \bibinfo {author}
  {\bibfnamefont {L.}~\bibnamefont {Zhang}}, \ and\ \bibinfo {author}
  {\bibfnamefont {Y.}~\bibnamefont {Ma}},\ }\href {\doibase
  10.1103/PhysRevX.7.011017} {\bibfield  {journal} {\bibinfo  {journal} {Phys.
  Rev. X}\ }\textbf {\bibinfo {volume} {7}},\ \bibinfo {pages} {011017}
  (\bibinfo {year} {2017})}\BibitemShut {NoStop}%
\bibitem [{\citenamefont {Wang}\ \emph {et~al.}(2017)\citenamefont {Wang},
  \citenamefont {Hanzawa}, \citenamefont {Hiramatsu}, \citenamefont {Kim},
  \citenamefont {Umezawa}, \citenamefont {Iwanaka}, \citenamefont {Tada},\ and\
  \citenamefont {Hosono}}]{wang2017exploration}%
  \BibitemOpen
  \bibfield  {author} {\bibinfo {author} {\bibfnamefont {J.}~\bibnamefont
  {Wang}}, \bibinfo {author} {\bibfnamefont {K.}~\bibnamefont {Hanzawa}},
  \bibinfo {author} {\bibfnamefont {H.}~\bibnamefont {Hiramatsu}}, \bibinfo
  {author} {\bibfnamefont {J.}~\bibnamefont {Kim}}, \bibinfo {author}
  {\bibfnamefont {N.}~\bibnamefont {Umezawa}}, \bibinfo {author} {\bibfnamefont
  {K.}~\bibnamefont {Iwanaka}}, \bibinfo {author} {\bibfnamefont
  {T.}~\bibnamefont {Tada}}, \ and\ \bibinfo {author} {\bibfnamefont
  {H.}~\bibnamefont {Hosono}},\ }\href {\doibase 10.1021/jacs.7b06279}
  {\bibfield  {journal} {\bibinfo  {journal} {Journal of the American Chemical
  Society}\ }\textbf {\bibinfo {volume} {139}},\ \bibinfo {pages} {15668}
  (\bibinfo {year} {2017})}\BibitemShut {NoStop}%
\bibitem [{\citenamefont {Zhu}\ \emph {et~al.}(2019{\natexlab{b}})\citenamefont
  {Zhu}, \citenamefont {Wang}, \citenamefont {Qu}, \citenamefont {Wang},
  \citenamefont {Frolov}, \citenamefont {Chen},\ and\ \citenamefont
  {Zhu}}]{PhysRevMaterials.3.024205}%
  \BibitemOpen
  \bibfield  {author} {\bibinfo {author} {\bibfnamefont {S.-C.}\ \bibnamefont
  {Zhu}}, \bibinfo {author} {\bibfnamefont {L.}~\bibnamefont {Wang}}, \bibinfo
  {author} {\bibfnamefont {J.-Y.}\ \bibnamefont {Qu}}, \bibinfo {author}
  {\bibfnamefont {J.-J.}\ \bibnamefont {Wang}}, \bibinfo {author}
  {\bibfnamefont {T.}~\bibnamefont {Frolov}}, \bibinfo {author} {\bibfnamefont
  {X.-Q.}\ \bibnamefont {Chen}}, \ and\ \bibinfo {author} {\bibfnamefont
  {Q.}~\bibnamefont {Zhu}},\ }\href {\doibase
  10.1103/PhysRevMaterials.3.024205} {\bibfield  {journal} {\bibinfo  {journal}
  {Phys. Rev. Materials}\ }\textbf {\bibinfo {volume} {3}},\ \bibinfo {pages}
  {024205} (\bibinfo {year} {2019}{\natexlab{b}})}\BibitemShut {NoStop}%
\bibitem [{\citenamefont {Oganov}\ \emph {et~al.}(2019)\citenamefont {Oganov},
  \citenamefont {Pickard}, \citenamefont {Zhu},\ and\ \citenamefont
  {Needs}}]{oganov2019structure}%
  \BibitemOpen
  \bibfield  {author} {\bibinfo {author} {\bibfnamefont {A.~R.}\ \bibnamefont
  {Oganov}}, \bibinfo {author} {\bibfnamefont {C.~J.}\ \bibnamefont {Pickard}},
  \bibinfo {author} {\bibfnamefont {Q.}~\bibnamefont {Zhu}}, \ and\ \bibinfo
  {author} {\bibfnamefont {R.~J.}\ \bibnamefont {Needs}},\ }\href
  {https://doi.org/10.1038/s41578-019-0101-8} {\bibfield  {journal} {\bibinfo
  {journal} {Nature Reviews Materials}\ }\textbf {\bibinfo {volume} {4}},\
  \bibinfo {pages} {331} (\bibinfo {year} {2019})}\BibitemShut {NoStop}%
\bibitem [{\citenamefont {Stevanovi\ifmmode~\acute{c}\else
  \'{c}\fi{}}(2016)}]{PhysRevLett.116.075503}%
  \BibitemOpen
  \bibfield  {author} {\bibinfo {author} {\bibfnamefont {V.}~\bibnamefont
  {Stevanovi\ifmmode~\acute{c}\else \'{c}\fi{}}},\ }\href {\doibase
  10.1103/PhysRevLett.116.075503} {\bibfield  {journal} {\bibinfo  {journal}
  {Phys. Rev. Lett.}\ }\textbf {\bibinfo {volume} {116}},\ \bibinfo {pages}
  {075503} (\bibinfo {year} {2016})}\BibitemShut {NoStop}%
\bibitem [{\citenamefont {Sun}\ \emph {et~al.}(2016)\citenamefont {Sun},
  \citenamefont {Dacek}, \citenamefont {Ong}, \citenamefont {Hautier},
  \citenamefont {Jain}, \citenamefont {Richards}, \citenamefont {Gamst},
  \citenamefont {Persson},\ and\ \citenamefont
  {Ceder}}]{sun1600225thermodynamic}%
  \BibitemOpen
  \bibfield  {author} {\bibinfo {author} {\bibfnamefont {W.}~\bibnamefont
  {Sun}}, \bibinfo {author} {\bibfnamefont {S.~T.}\ \bibnamefont {Dacek}},
  \bibinfo {author} {\bibfnamefont {S.~P.}\ \bibnamefont {Ong}}, \bibinfo
  {author} {\bibfnamefont {G.}~\bibnamefont {Hautier}}, \bibinfo {author}
  {\bibfnamefont {A.}~\bibnamefont {Jain}}, \bibinfo {author} {\bibfnamefont
  {W.~D.}\ \bibnamefont {Richards}}, \bibinfo {author} {\bibfnamefont {A.~C.}\
  \bibnamefont {Gamst}}, \bibinfo {author} {\bibfnamefont {K.~A.}\ \bibnamefont
  {Persson}}, \ and\ \bibinfo {author} {\bibfnamefont {G.}~\bibnamefont
  {Ceder}},\ }\href {https://advances.sciencemag.org/content/2/11/e1600225}
  {\bibfield  {journal} {\bibinfo  {journal} {Science Advances}\ }\textbf
  {\bibinfo {volume} {2}} (\bibinfo {year} {2016})}\BibitemShut {NoStop}%
\bibitem [{\citenamefont {Kresse}\ and\ \citenamefont
  {Joubert}(1999)}]{PhysRevB.59.1758}%
  \BibitemOpen
  \bibfield  {author} {\bibinfo {author} {\bibfnamefont {G.}~\bibnamefont
  {Kresse}}\ and\ \bibinfo {author} {\bibfnamefont {D.}~\bibnamefont
  {Joubert}},\ }\href {\doibase 10.1103/PhysRevB.59.1758} {\bibfield  {journal}
  {\bibinfo  {journal} {Phys. Rev. B}\ }\textbf {\bibinfo {volume} {59}},\
  \bibinfo {pages} {1758} (\bibinfo {year} {1999})}\BibitemShut {NoStop}%
\bibitem [{\citenamefont {Kresse}\ and\ \citenamefont
  {Furthm\"uller}(1996)}]{PhysRevB.54.11169}%
  \BibitemOpen
  \bibfield  {author} {\bibinfo {author} {\bibfnamefont {G.}~\bibnamefont
  {Kresse}}\ and\ \bibinfo {author} {\bibfnamefont {J.}~\bibnamefont
  {Furthm\"uller}},\ }\href {\doibase 10.1103/PhysRevB.54.11169} {\bibfield
  {journal} {\bibinfo  {journal} {Phys. Rev. B}\ }\textbf {\bibinfo {volume}
  {54}},\ \bibinfo {pages} {11169} (\bibinfo {year} {1996})}\BibitemShut
  {NoStop}%
\bibitem [{\citenamefont {Perdew}\ \emph {et~al.}(1996)\citenamefont {Perdew},
  \citenamefont {Burke},\ and\ \citenamefont
  {Ernzerhof}}]{PhysRevLett.77.3865}%
  \BibitemOpen
  \bibfield  {author} {\bibinfo {author} {\bibfnamefont {J.~P.}\ \bibnamefont
  {Perdew}}, \bibinfo {author} {\bibfnamefont {K.}~\bibnamefont {Burke}}, \
  and\ \bibinfo {author} {\bibfnamefont {M.}~\bibnamefont {Ernzerhof}},\ }\href
  {\doibase 10.1103/PhysRevLett.77.3865} {\bibfield  {journal} {\bibinfo
  {journal} {Phys. Rev. Lett.}\ }\textbf {\bibinfo {volume} {77}},\ \bibinfo
  {pages} {3865} (\bibinfo {year} {1996})}\BibitemShut {NoStop}%
\bibitem [{\citenamefont {Liechtenstein}\ \emph {et~al.}(1995)\citenamefont
  {Liechtenstein}, \citenamefont {Anisimov},\ and\ \citenamefont
  {Zaanen}}]{PhysRevB.52.R5467}%
  \BibitemOpen
  \bibfield  {author} {\bibinfo {author} {\bibfnamefont {A.~I.}\ \bibnamefont
  {Liechtenstein}}, \bibinfo {author} {\bibfnamefont {V.~I.}\ \bibnamefont
  {Anisimov}}, \ and\ \bibinfo {author} {\bibfnamefont {J.}~\bibnamefont
  {Zaanen}},\ }\href {\doibase 10.1103/PhysRevB.52.R5467} {\bibfield  {journal}
  {\bibinfo  {journal} {Phys. Rev. B}\ }\textbf {\bibinfo {volume} {52}},\
  \bibinfo {pages} {R5467} (\bibinfo {year} {1995})}\BibitemShut {NoStop}%
\bibitem [{\citenamefont {Heyd}\ \emph {et~al.}(2003)\citenamefont {Heyd},
  \citenamefont {Scuseria},\ and\ \citenamefont {Ernzerhof}}]{heyd2003hybrid}%
  \BibitemOpen
  \bibfield  {author} {\bibinfo {author} {\bibfnamefont {J.}~\bibnamefont
  {Heyd}}, \bibinfo {author} {\bibfnamefont {G.~E.}\ \bibnamefont {Scuseria}},
  \ and\ \bibinfo {author} {\bibfnamefont {M.}~\bibnamefont {Ernzerhof}},\
  }\href {\doibase 10.1063/1.1564060} {\bibfield  {journal} {\bibinfo
  {journal} {The Journal of chemical physics}\ }\textbf {\bibinfo {volume}
  {118}},\ \bibinfo {pages} {8207} (\bibinfo {year} {2003})}\BibitemShut
  {NoStop}%
\bibitem [{\citenamefont {Wang}\ \emph
  {et~al.}(2019{\natexlab{b}})\citenamefont {Wang}, \citenamefont {Xu},
  \citenamefont {Liu}, \citenamefont {Tang},\ and\ \citenamefont
  {Geng}}]{wang2019vaspkit}%
  \BibitemOpen
  \bibfield  {author} {\bibinfo {author} {\bibfnamefont {V.}~\bibnamefont
  {Wang}}, \bibinfo {author} {\bibfnamefont {N.}~\bibnamefont {Xu}}, \bibinfo
  {author} {\bibfnamefont {J.}~\bibnamefont {Liu}}, \bibinfo {author}
  {\bibfnamefont {G.}~\bibnamefont {Tang}}, \ and\ \bibinfo {author}
  {\bibfnamefont {W.}~\bibnamefont {Geng}},\ }\href
  {https://arxiv.org/abs/1908.08269} {\bibfield  {journal} {\bibinfo  {journal}
  {arXiv preprint arXiv:1908.08269}\ } (\bibinfo {year}
  {2019}{\natexlab{b}})}\BibitemShut {NoStop}%
\bibitem [{\citenamefont {Herath}\ \emph {et~al.}(2020)\citenamefont {Herath},
  \citenamefont {Tavadze}, \citenamefont {He}, \citenamefont {Bousquet},
  \citenamefont {Singh}, \citenamefont {Muñoz},\ and\ \citenamefont
  {Romero}}]{herath2020107080}%
  \BibitemOpen
  \bibfield  {author} {\bibinfo {author} {\bibfnamefont {U.}~\bibnamefont
  {Herath}}, \bibinfo {author} {\bibfnamefont {P.}~\bibnamefont {Tavadze}},
  \bibinfo {author} {\bibfnamefont {X.}~\bibnamefont {He}}, \bibinfo {author}
  {\bibfnamefont {E.}~\bibnamefont {Bousquet}}, \bibinfo {author}
  {\bibfnamefont {S.}~\bibnamefont {Singh}}, \bibinfo {author} {\bibfnamefont
  {F.}~\bibnamefont {Muñoz}}, \ and\ \bibinfo {author} {\bibfnamefont {A.~H.}\
  \bibnamefont {Romero}},\ }\href {\doibase
  https://doi.org/10.1016/j.cpc.2019.107080} {\bibfield  {journal} {\bibinfo
  {journal} {Computer Physics Communications}\ }\textbf {\bibinfo {volume}
  {251}},\ \bibinfo {pages} {107080} (\bibinfo {year} {2020})}\BibitemShut
  {NoStop}%
\bibitem [{\citenamefont {Setyawan}\ and\ \citenamefont
  {Curtarolo}(2010{\natexlab{a}})}]{setyawan2010299}%
  \BibitemOpen
  \bibfield  {author} {\bibinfo {author} {\bibfnamefont {W.}~\bibnamefont
  {Setyawan}}\ and\ \bibinfo {author} {\bibfnamefont {S.}~\bibnamefont
  {Curtarolo}},\ }\href {\doibase
  https://doi.org/10.1016/j.commatsci.2010.05.010} {\bibfield  {journal}
  {\bibinfo  {journal} {Computational Materials Science}\ }\textbf {\bibinfo
  {volume} {49}},\ \bibinfo {pages} {299 } (\bibinfo {year}
  {2010}{\natexlab{a}})}\BibitemShut {NoStop}%
\bibitem [{\citenamefont {Togo}\ and\ \citenamefont
  {Tanaka}(2018)}]{togo2018texttt}%
  \BibitemOpen
  \bibfield  {author} {\bibinfo {author} {\bibfnamefont {A.}~\bibnamefont
  {Togo}}\ and\ \bibinfo {author} {\bibfnamefont {I.}~\bibnamefont {Tanaka}},\
  }\href {https://arxiv.org/abs/1808.01590} {\bibfield  {journal} {\bibinfo
  {journal} {arXiv preprint arXiv:1808.01590}\ } (\bibinfo {year}
  {2018})}\BibitemShut {NoStop}%
\bibitem [{\citenamefont {Momma}\ and\ \citenamefont
  {Izumi}(2008)}]{momma2008vesta}%
  \BibitemOpen
  \bibfield  {author} {\bibinfo {author} {\bibfnamefont {K.}~\bibnamefont
  {Momma}}\ and\ \bibinfo {author} {\bibfnamefont {F.}~\bibnamefont {Izumi}},\
  }\href {\doibase 10.1107/S0021889808012016} {\bibfield  {journal} {\bibinfo
  {journal} {Journal of Applied Crystallography}\ }\textbf {\bibinfo {volume}
  {41}},\ \bibinfo {pages} {653} (\bibinfo {year} {2008})}\BibitemShut
  {NoStop}%
\bibitem [{\citenamefont {Goedecker}(2004)}]{goedecker2004minima}%
  \BibitemOpen
  \bibfield  {author} {\bibinfo {author} {\bibfnamefont {S.}~\bibnamefont
  {Goedecker}},\ }\href {\doibase 10.1063/1.1724816} {\bibfield  {journal}
  {\bibinfo  {journal} {The Journal of Chemical Physics}\ }\textbf {\bibinfo
  {volume} {120}},\ \bibinfo {pages} {9911} (\bibinfo {year}
  {2004})}\BibitemShut {NoStop}%
\bibitem [{\citenamefont {Amsler}\ and\ \citenamefont
  {Goedecker}(2010)}]{amsler2010crystal}%
  \BibitemOpen
  \bibfield  {author} {\bibinfo {author} {\bibfnamefont {M.}~\bibnamefont
  {Amsler}}\ and\ \bibinfo {author} {\bibfnamefont {S.}~\bibnamefont
  {Goedecker}},\ }\href {\doibase 10.1063/1.3512900} {\bibfield  {journal}
  {\bibinfo  {journal} {The Journal of Chemical Physics}\ }\textbf {\bibinfo
  {volume} {133}},\ \bibinfo {pages} {224104} (\bibinfo {year}
  {2010})}\BibitemShut {NoStop}%
\bibitem [{\citenamefont {Flores-Livas}(2020)}]{flores2020crystal}%
  \BibitemOpen
  \bibfield  {author} {\bibinfo {author} {\bibfnamefont {J.~A.}\ \bibnamefont
  {Flores-Livas}},\ }\href {\doibase 10.1088/1361-648x/ab7e54} {\bibfield
  {journal} {\bibinfo  {journal} {Journal of Physics: Condensed Matter}\
  }\textbf {\bibinfo {volume} {32}},\ \bibinfo {pages} {294002} (\bibinfo
  {year} {2020})}\BibitemShut {NoStop}%
\bibitem [{\citenamefont {Flores-Livas}\ \emph {et~al.}(2020)\citenamefont
  {Flores-Livas}, \citenamefont {Boeri}, \citenamefont {Sanna}, \citenamefont
  {Profeta}, \citenamefont {Arita},\ and\ \citenamefont
  {Eremets}}]{flores2020perspective}%
  \BibitemOpen
  \bibfield  {author} {\bibinfo {author} {\bibfnamefont {J.~A.}\ \bibnamefont
  {Flores-Livas}}, \bibinfo {author} {\bibfnamefont {L.}~\bibnamefont {Boeri}},
  \bibinfo {author} {\bibfnamefont {A.}~\bibnamefont {Sanna}}, \bibinfo
  {author} {\bibfnamefont {G.}~\bibnamefont {Profeta}}, \bibinfo {author}
  {\bibfnamefont {R.}~\bibnamefont {Arita}}, \ and\ \bibinfo {author}
  {\bibfnamefont {M.}~\bibnamefont {Eremets}},\ }\href {\doibase
  https://doi.org/10.1016/j.physrep.2020.02.003} {\bibfield  {journal}
  {\bibinfo  {journal} {Physics Reports}\ }\textbf {\bibinfo {volume} {856}},\
  \bibinfo {pages} {1 } (\bibinfo {year} {2020})}\BibitemShut {NoStop}%
\bibitem [{\citenamefont {Jain}\ \emph {et~al.}(2013)\citenamefont {Jain},
  \citenamefont {Ong}, \citenamefont {Hautier}, \citenamefont {Chen},
  \citenamefont {Richards}, \citenamefont {Dacek}, \citenamefont {Cholia},
  \citenamefont {Gunter}, \citenamefont {Skinner}, \citenamefont {Ceder},\ and\
  \citenamefont {Persson}}]{Jain2013Materials}%
  \BibitemOpen
  \bibfield  {author} {\bibinfo {author} {\bibfnamefont {A.}~\bibnamefont
  {Jain}}, \bibinfo {author} {\bibfnamefont {S.~P.}\ \bibnamefont {Ong}},
  \bibinfo {author} {\bibfnamefont {G.}~\bibnamefont {Hautier}}, \bibinfo
  {author} {\bibfnamefont {W.}~\bibnamefont {Chen}}, \bibinfo {author}
  {\bibfnamefont {W.~D.}\ \bibnamefont {Richards}}, \bibinfo {author}
  {\bibfnamefont {S.}~\bibnamefont {Dacek}}, \bibinfo {author} {\bibfnamefont
  {S.}~\bibnamefont {Cholia}}, \bibinfo {author} {\bibfnamefont
  {D.}~\bibnamefont {Gunter}}, \bibinfo {author} {\bibfnamefont
  {D.}~\bibnamefont {Skinner}}, \bibinfo {author} {\bibfnamefont
  {G.}~\bibnamefont {Ceder}}, \ and\ \bibinfo {author} {\bibfnamefont {K.~a.}\
  \bibnamefont {Persson}},\ }\href {\doibase 10.1063/1.4812323} {\bibfield
  {journal} {\bibinfo  {journal} {APL Materials}\ }\textbf {\bibinfo {volume}
  {1}},\ \bibinfo {pages} {011002} (\bibinfo {year} {2013})}\BibitemShut
  {NoStop}%
\bibitem [{\citenamefont {Ong}\ \emph {et~al.}(2008)\citenamefont {Ong},
  \citenamefont {Wang}, \citenamefont {Kang},\ and\ \citenamefont
  {Ceder}}]{Ong2008Li}%
  \BibitemOpen
  \bibfield  {author} {\bibinfo {author} {\bibfnamefont {S.~P.}\ \bibnamefont
  {Ong}}, \bibinfo {author} {\bibfnamefont {L.}~\bibnamefont {Wang}}, \bibinfo
  {author} {\bibfnamefont {B.}~\bibnamefont {Kang}}, \ and\ \bibinfo {author}
  {\bibfnamefont {G.}~\bibnamefont {Ceder}},\ }\href {\doibase
  10.1021/cm702327g} {\bibfield  {journal} {\bibinfo  {journal} {Chemistry of
  Materials}\ }\textbf {\bibinfo {volume} {20}},\ \bibinfo {pages} {1798}
  (\bibinfo {year} {2008})}\BibitemShut {NoStop}%
\bibitem [{\citenamefont {Togo}\ and\ \citenamefont
  {Tanaka}(2015)}]{togo2015first}%
  \BibitemOpen
  \bibfield  {author} {\bibinfo {author} {\bibfnamefont {A.}~\bibnamefont
  {Togo}}\ and\ \bibinfo {author} {\bibfnamefont {I.}~\bibnamefont {Tanaka}},\
  }\href {\doibase https://doi.org/10.1016/j.scriptamat.2015.07.021} {\bibfield
   {journal} {\bibinfo  {journal} {Scripta Materialia}\ }\textbf {\bibinfo
  {volume} {108}},\ \bibinfo {pages} {1 } (\bibinfo {year} {2015})}\BibitemShut
  {NoStop}%
\bibitem [{\citenamefont {Huebsch}\ \emph {et~al.}(2020)\citenamefont
  {Huebsch}, \citenamefont {Nomoto}, \citenamefont {Suzuki},\ and\
  \citenamefont {Arita}}]{huebsch2020benchmark}%
  \BibitemOpen
  \bibfield  {author} {\bibinfo {author} {\bibfnamefont {M.-T.}\ \bibnamefont
  {Huebsch}}, \bibinfo {author} {\bibfnamefont {T.}~\bibnamefont {Nomoto}},
  \bibinfo {author} {\bibfnamefont {M.-T.}\ \bibnamefont {Suzuki}}, \ and\
  \bibinfo {author} {\bibfnamefont {R.}~\bibnamefont {Arita}},\ }\href
  {https://arxiv.org/abs/2008.13669} {\bibfield  {journal} {\bibinfo  {journal}
  {arXiv preprint arXiv:2008.13669}\ } (\bibinfo {year} {2020})}\BibitemShut
  {NoStop}%
\bibitem [{\citenamefont {Suzuki}\ \emph {et~al.}(2017)\citenamefont {Suzuki},
  \citenamefont {Koretsune}, \citenamefont {Ochi},\ and\ \citenamefont
  {Arita}}]{PhysRevB.95.094406}%
  \BibitemOpen
  \bibfield  {author} {\bibinfo {author} {\bibfnamefont {M.-T.}\ \bibnamefont
  {Suzuki}}, \bibinfo {author} {\bibfnamefont {T.}~\bibnamefont {Koretsune}},
  \bibinfo {author} {\bibfnamefont {M.}~\bibnamefont {Ochi}}, \ and\ \bibinfo
  {author} {\bibfnamefont {R.}~\bibnamefont {Arita}},\ }\href {\doibase
  10.1103/PhysRevB.95.094406} {\bibfield  {journal} {\bibinfo  {journal} {Phys.
  Rev. B}\ }\textbf {\bibinfo {volume} {95}},\ \bibinfo {pages} {094406}
  (\bibinfo {year} {2017})}\BibitemShut {NoStop}%
\bibitem [{\citenamefont {Suzuki}\ \emph {et~al.}(2019)\citenamefont {Suzuki},
  \citenamefont {Nomoto}, \citenamefont {Arita}, \citenamefont {Yanagi},
  \citenamefont {Hayami},\ and\ \citenamefont {Kusunose}}]{PhysRevB.99.174407}%
  \BibitemOpen
  \bibfield  {author} {\bibinfo {author} {\bibfnamefont {M.-T.}\ \bibnamefont
  {Suzuki}}, \bibinfo {author} {\bibfnamefont {T.}~\bibnamefont {Nomoto}},
  \bibinfo {author} {\bibfnamefont {R.}~\bibnamefont {Arita}}, \bibinfo
  {author} {\bibfnamefont {Y.}~\bibnamefont {Yanagi}}, \bibinfo {author}
  {\bibfnamefont {S.}~\bibnamefont {Hayami}}, \ and\ \bibinfo {author}
  {\bibfnamefont {H.}~\bibnamefont {Kusunose}},\ }\href {\doibase
  10.1103/PhysRevB.99.174407} {\bibfield  {journal} {\bibinfo  {journal} {Phys.
  Rev. B}\ }\textbf {\bibinfo {volume} {99}},\ \bibinfo {pages} {174407}
  (\bibinfo {year} {2019})}\BibitemShut {NoStop}%
\bibitem [{\citenamefont {Gao}\ \emph {et~al.}(2020)\citenamefont {Gao},
  \citenamefont {Wu}, \citenamefont {Persson},\ and\ \citenamefont
  {Wang}}]{gao2020irvsp}%
  \BibitemOpen
  \bibfield  {author} {\bibinfo {author} {\bibfnamefont {J.}~\bibnamefont
  {Gao}}, \bibinfo {author} {\bibfnamefont {Q.}~\bibnamefont {Wu}}, \bibinfo
  {author} {\bibfnamefont {C.}~\bibnamefont {Persson}}, \ and\ \bibinfo
  {author} {\bibfnamefont {Z.}~\bibnamefont {Wang}},\ }\href
  {https://arxiv.org/abs/2002.04032} {\bibfield  {journal} {\bibinfo  {journal}
  {arXiv preprint arXiv:2002.04032}\ } (\bibinfo {year} {2020})}\BibitemShut
  {NoStop}%
\bibitem [{\citenamefont {Wu}\ \emph {et~al.}(2018)\citenamefont {Wu},
  \citenamefont {Zhang}, \citenamefont {Song}, \citenamefont {Troyer},\ and\
  \citenamefont {Soluyanov}}]{wu2018wanniertools}%
  \BibitemOpen
  \bibfield  {author} {\bibinfo {author} {\bibfnamefont {Q.}~\bibnamefont
  {Wu}}, \bibinfo {author} {\bibfnamefont {S.}~\bibnamefont {Zhang}}, \bibinfo
  {author} {\bibfnamefont {H.-F.}\ \bibnamefont {Song}}, \bibinfo {author}
  {\bibfnamefont {M.}~\bibnamefont {Troyer}}, \ and\ \bibinfo {author}
  {\bibfnamefont {A.~A.}\ \bibnamefont {Soluyanov}},\ }\href {\doibase
  https://doi.org/10.1016/j.cpc.2017.09.033} {\bibfield  {journal} {\bibinfo
  {journal} {Computer Physics Communications}\ }\textbf {\bibinfo {volume}
  {224}},\ \bibinfo {pages} {405 } (\bibinfo {year} {2018})}\BibitemShut
  {NoStop}%
\bibitem [{\citenamefont {Marzari}\ and\ \citenamefont
  {Vanderbilt}(1997)}]{PhysRevB.56.12847}%
  \BibitemOpen
  \bibfield  {author} {\bibinfo {author} {\bibfnamefont {N.}~\bibnamefont
  {Marzari}}\ and\ \bibinfo {author} {\bibfnamefont {D.}~\bibnamefont
  {Vanderbilt}},\ }\href {\doibase 10.1103/PhysRevB.56.12847} {\bibfield
  {journal} {\bibinfo  {journal} {Phys. Rev. B}\ }\textbf {\bibinfo {volume}
  {56}},\ \bibinfo {pages} {12847} (\bibinfo {year} {1997})}\BibitemShut
  {NoStop}%
\bibitem [{\citenamefont {Souza}\ \emph {et~al.}(2001)\citenamefont {Souza},
  \citenamefont {Marzari},\ and\ \citenamefont
  {Vanderbilt}}]{PhysRevB.65.035109}%
  \BibitemOpen
  \bibfield  {author} {\bibinfo {author} {\bibfnamefont {I.}~\bibnamefont
  {Souza}}, \bibinfo {author} {\bibfnamefont {N.}~\bibnamefont {Marzari}}, \
  and\ \bibinfo {author} {\bibfnamefont {D.}~\bibnamefont {Vanderbilt}},\
  }\href {\doibase 10.1103/PhysRevB.65.035109} {\bibfield  {journal} {\bibinfo
  {journal} {Phys. Rev. B}\ }\textbf {\bibinfo {volume} {65}},\ \bibinfo
  {pages} {035109} (\bibinfo {year} {2001})}\BibitemShut {NoStop}%
\bibitem [{\citenamefont {Pizzi}\ \emph {et~al.}(2020)\citenamefont {Pizzi},
  \citenamefont {Vitale}, \citenamefont {Arita}, \citenamefont {Blügel},
  \citenamefont {Freimuth}, \citenamefont {G{\'{e}}ranton}, \citenamefont
  {Gibertini}, \citenamefont {Gresch}, \citenamefont {Johnson}, \citenamefont
  {Koretsune}, \citenamefont {Iba{\~{n}}ez-Azpiroz}, \citenamefont {Lee},
  \citenamefont {Lihm}, \citenamefont {Marchand}, \citenamefont {Marrazzo},
  \citenamefont {Mokrousov}, \citenamefont {Mustafa}, \citenamefont {Nohara},
  \citenamefont {Nomura}, \citenamefont {Paulatto}, \citenamefont
  {Ponc{\'{e}}}, \citenamefont {Ponweiser}, \citenamefont {Qiao}, \citenamefont
  {Thöle}, \citenamefont {Tsirkin}, \citenamefont {Wierzbowska}, \citenamefont
  {Marzari}, \citenamefont {Vanderbilt}, \citenamefont {Souza}, \citenamefont
  {Mostofi},\ and\ \citenamefont {Yates}}]{pizzi2020wannier90}%
  \BibitemOpen
  \bibfield  {author} {\bibinfo {author} {\bibfnamefont {G.}~\bibnamefont
  {Pizzi}}, \bibinfo {author} {\bibfnamefont {V.}~\bibnamefont {Vitale}},
  \bibinfo {author} {\bibfnamefont {R.}~\bibnamefont {Arita}}, \bibinfo
  {author} {\bibfnamefont {S.}~\bibnamefont {Blügel}}, \bibinfo {author}
  {\bibfnamefont {F.}~\bibnamefont {Freimuth}}, \bibinfo {author}
  {\bibfnamefont {G.}~\bibnamefont {G{\'{e}}ranton}}, \bibinfo {author}
  {\bibfnamefont {M.}~\bibnamefont {Gibertini}}, \bibinfo {author}
  {\bibfnamefont {D.}~\bibnamefont {Gresch}}, \bibinfo {author} {\bibfnamefont
  {C.}~\bibnamefont {Johnson}}, \bibinfo {author} {\bibfnamefont
  {T.}~\bibnamefont {Koretsune}}, \bibinfo {author} {\bibfnamefont
  {J.}~\bibnamefont {Iba{\~{n}}ez-Azpiroz}}, \bibinfo {author} {\bibfnamefont
  {H.}~\bibnamefont {Lee}}, \bibinfo {author} {\bibfnamefont {J.-M.}\
  \bibnamefont {Lihm}}, \bibinfo {author} {\bibfnamefont {D.}~\bibnamefont
  {Marchand}}, \bibinfo {author} {\bibfnamefont {A.}~\bibnamefont {Marrazzo}},
  \bibinfo {author} {\bibfnamefont {Y.}~\bibnamefont {Mokrousov}}, \bibinfo
  {author} {\bibfnamefont {J.~I.}\ \bibnamefont {Mustafa}}, \bibinfo {author}
  {\bibfnamefont {Y.}~\bibnamefont {Nohara}}, \bibinfo {author} {\bibfnamefont
  {Y.}~\bibnamefont {Nomura}}, \bibinfo {author} {\bibfnamefont
  {L.}~\bibnamefont {Paulatto}}, \bibinfo {author} {\bibfnamefont
  {S.}~\bibnamefont {Ponc{\'{e}}}}, \bibinfo {author} {\bibfnamefont
  {T.}~\bibnamefont {Ponweiser}}, \bibinfo {author} {\bibfnamefont
  {J.}~\bibnamefont {Qiao}}, \bibinfo {author} {\bibfnamefont {F.}~\bibnamefont
  {Thöle}}, \bibinfo {author} {\bibfnamefont {S.~S.}\ \bibnamefont {Tsirkin}},
  \bibinfo {author} {\bibfnamefont {M.}~\bibnamefont {Wierzbowska}}, \bibinfo
  {author} {\bibfnamefont {N.}~\bibnamefont {Marzari}}, \bibinfo {author}
  {\bibfnamefont {D.}~\bibnamefont {Vanderbilt}}, \bibinfo {author}
  {\bibfnamefont {I.}~\bibnamefont {Souza}}, \bibinfo {author} {\bibfnamefont
  {A.~A.}\ \bibnamefont {Mostofi}}, \ and\ \bibinfo {author} {\bibfnamefont
  {J.~R.}\ \bibnamefont {Yates}},\ }\href {\doibase 10.1088/1361-648x/ab51ff}
  {\bibfield  {journal} {\bibinfo  {journal} {Journal of Physics: Condensed
  Matter}\ }\textbf {\bibinfo {volume} {32}},\ \bibinfo {pages} {165902}
  (\bibinfo {year} {2020})}\BibitemShut {NoStop}%
\bibitem [{\citenamefont {Becke}\ and\ \citenamefont
  {Edgecombe}(1990)}]{becke1990simple}%
  \BibitemOpen
  \bibfield  {author} {\bibinfo {author} {\bibfnamefont {A.~D.}\ \bibnamefont
  {Becke}}\ and\ \bibinfo {author} {\bibfnamefont {K.~E.}\ \bibnamefont
  {Edgecombe}},\ }\href {\doibase 10.1063/1.458517} {\bibfield  {journal}
  {\bibinfo  {journal} {The Journal of Chemical Physics}\ }\textbf {\bibinfo
  {volume} {92}},\ \bibinfo {pages} {5397} (\bibinfo {year}
  {1990})}\BibitemShut {NoStop}%
\bibitem [{\citenamefont {Savin}(2005)}]{savin2005electron}%
  \BibitemOpen
  \bibfield  {author} {\bibinfo {author} {\bibfnamefont {A.}~\bibnamefont
  {Savin}},\ }\href {\doibase https://doi.org/10.1016/j.theochem.2005.02.034}
  {\bibfield  {journal} {\bibinfo  {journal} {Journal of Molecular Structure:
  THEOCHEM}\ }\textbf {\bibinfo {volume} {727}},\ \bibinfo {pages} {127 }
  (\bibinfo {year} {2005})}\BibitemShut {NoStop}%
\bibitem [{\citenamefont {Zhao}\ \emph {et~al.}(2016)\citenamefont {Zhao},
  \citenamefont {Kan},\ and\ \citenamefont {Li}}]{zhao2016electride}%
  \BibitemOpen
  \bibfield  {author} {\bibinfo {author} {\bibfnamefont {S.}~\bibnamefont
  {Zhao}}, \bibinfo {author} {\bibfnamefont {E.}~\bibnamefont {Kan}}, \ and\
  \bibinfo {author} {\bibfnamefont {Z.}~\bibnamefont {Li}},\ }\href {\doibase
  10.1002/wcms.1258} {\bibfield  {journal} {\bibinfo  {journal} {WIREs
  Computational Molecular Science}\ }\textbf {\bibinfo {volume} {6}},\ \bibinfo
  {pages} {430} (\bibinfo {year} {2016})}\BibitemShut {NoStop}%
\bibitem [{\citenamefont {Dale}\ and\ \citenamefont
  {Johnson}(2018)}]{dale2018theoretical}%
  \BibitemOpen
  \bibfield  {author} {\bibinfo {author} {\bibfnamefont {S.~G.}\ \bibnamefont
  {Dale}}\ and\ \bibinfo {author} {\bibfnamefont {E.~R.}\ \bibnamefont
  {Johnson}},\ }\href {\doibase 10.1021/acs.jpca.8b08548} {\bibfield  {journal}
  {\bibinfo  {journal} {The Journal of Physical Chemistry A}\ }\textbf
  {\bibinfo {volume} {122}},\ \bibinfo {pages} {9371} (\bibinfo {year}
  {2018})}\BibitemShut {NoStop}%
\bibitem [{\citenamefont {Bergerhoff}\ \emph {et~al.}(1983)\citenamefont
  {Bergerhoff}, \citenamefont {Hundt}, \citenamefont {Sievers},\ and\
  \citenamefont {Brown}}]{bergerhoff1983inorganic}%
  \BibitemOpen
  \bibfield  {author} {\bibinfo {author} {\bibfnamefont {G.}~\bibnamefont
  {Bergerhoff}}, \bibinfo {author} {\bibfnamefont {R.}~\bibnamefont {Hundt}},
  \bibinfo {author} {\bibfnamefont {R.}~\bibnamefont {Sievers}}, \ and\
  \bibinfo {author} {\bibfnamefont {I.~D.}\ \bibnamefont {Brown}},\ }\href
  {\doibase 10.1021/ci00038a003} {\bibfield  {journal} {\bibinfo  {journal}
  {Journal of Chemical Information and Computer Sciences}\ }\textbf {\bibinfo
  {volume} {23}},\ \bibinfo {pages} {66} (\bibinfo {year} {1983})}\BibitemShut
  {NoStop}%
\bibitem [{\citenamefont {Zhang}\ \emph {et~al.}(2014)\citenamefont {Zhang},
  \citenamefont {Xiao}, \citenamefont {Lei}, \citenamefont {Toda},
  \citenamefont {Matsuishi}, \citenamefont {Kamiya}, \citenamefont {Ueda},\
  and\ \citenamefont {Hosono}}]{zhang2014two}%
  \BibitemOpen
  \bibfield  {author} {\bibinfo {author} {\bibfnamefont {X.}~\bibnamefont
  {Zhang}}, \bibinfo {author} {\bibfnamefont {Z.}~\bibnamefont {Xiao}},
  \bibinfo {author} {\bibfnamefont {H.}~\bibnamefont {Lei}}, \bibinfo {author}
  {\bibfnamefont {Y.}~\bibnamefont {Toda}}, \bibinfo {author} {\bibfnamefont
  {S.}~\bibnamefont {Matsuishi}}, \bibinfo {author} {\bibfnamefont
  {T.}~\bibnamefont {Kamiya}}, \bibinfo {author} {\bibfnamefont
  {S.}~\bibnamefont {Ueda}}, \ and\ \bibinfo {author} {\bibfnamefont
  {H.}~\bibnamefont {Hosono}},\ }\href {\doibase 10.1021/cm503512h} {\bibfield
  {journal} {\bibinfo  {journal} {Chemistry of Materials}\ }\textbf {\bibinfo
  {volume} {26}},\ \bibinfo {pages} {6638} (\bibinfo {year}
  {2014})}\BibitemShut {NoStop}%
\bibitem [{\citenamefont {Lee}\ \emph {et~al.}(2020)\citenamefont {Lee},
  \citenamefont {Hwang}, \citenamefont {Park}, \citenamefont {Nandadasa},
  \citenamefont {Kim}, \citenamefont {Bang}, \citenamefont {Lee}, \citenamefont
  {Lee}, \citenamefont {Zhang}, \citenamefont {Ma}, \citenamefont {Hosono},
  \citenamefont {Lee}, \citenamefont {Kim},\ and\ \citenamefont
  {Kim}}]{lee2020ferromagnetic}%
  \BibitemOpen
  \bibfield  {author} {\bibinfo {author} {\bibfnamefont {S.~Y.}\ \bibnamefont
  {Lee}}, \bibinfo {author} {\bibfnamefont {J.-Y.}\ \bibnamefont {Hwang}},
  \bibinfo {author} {\bibfnamefont {J.}~\bibnamefont {Park}}, \bibinfo {author}
  {\bibfnamefont {C.~N.}\ \bibnamefont {Nandadasa}}, \bibinfo {author}
  {\bibfnamefont {Y.}~\bibnamefont {Kim}}, \bibinfo {author} {\bibfnamefont
  {J.}~\bibnamefont {Bang}}, \bibinfo {author} {\bibfnamefont {K.}~\bibnamefont
  {Lee}}, \bibinfo {author} {\bibfnamefont {K.~H.}\ \bibnamefont {Lee}},
  \bibinfo {author} {\bibfnamefont {Y.}~\bibnamefont {Zhang}}, \bibinfo
  {author} {\bibfnamefont {Y.}~\bibnamefont {Ma}}, \bibinfo {author}
  {\bibfnamefont {H.}~\bibnamefont {Hosono}}, \bibinfo {author} {\bibfnamefont
  {Y.~H.}\ \bibnamefont {Lee}}, \bibinfo {author} {\bibfnamefont {S.-G.}\
  \bibnamefont {Kim}}, \ and\ \bibinfo {author} {\bibfnamefont {S.~W.}\
  \bibnamefont {Kim}},\ }\href
  {https://www.nature.com/articles/s41467-020-15253-5} {\bibfield  {journal}
  {\bibinfo  {journal} {Nature communications}\ }\textbf {\bibinfo {volume}
  {11}},\ \bibinfo {pages} {1} (\bibinfo {year} {2020})}\BibitemShut {NoStop}%
\bibitem [{\citenamefont {Simon}\ \emph {et~al.}(1996)\citenamefont {Simon},
  \citenamefont {B{\"a}cker}, \citenamefont {Henn}, \citenamefont {Felser},
  \citenamefont {Kremer}, \citenamefont {Mattausch},\ and\ \citenamefont
  {Yoshiasa}}]{simon1996supraleitung}%
  \BibitemOpen
  \bibfield  {author} {\bibinfo {author} {\bibfnamefont {A.}~\bibnamefont
  {Simon}}, \bibinfo {author} {\bibfnamefont {M.}~\bibnamefont {B{\"a}cker}},
  \bibinfo {author} {\bibfnamefont {R.}~\bibnamefont {Henn}}, \bibinfo {author}
  {\bibfnamefont {C.}~\bibnamefont {Felser}}, \bibinfo {author} {\bibfnamefont
  {R.}~\bibnamefont {Kremer}}, \bibinfo {author} {\bibfnamefont
  {H.}~\bibnamefont {Mattausch}}, \ and\ \bibinfo {author} {\bibfnamefont
  {A.}~\bibnamefont {Yoshiasa}},\ }\href {\doibase 10.1002/zaac.19966220118}
  {\bibfield  {journal} {\bibinfo  {journal} {Zeitschrift f{\"u}r anorganische
  und allgemeine Chemie}\ }\textbf {\bibinfo {volume} {622}},\ \bibinfo {pages}
  {123} (\bibinfo {year} {1996})}\BibitemShut {NoStop}%
\bibitem [{\citenamefont {Chan}\ \emph {et~al.}(2016)\citenamefont {Chan},
  \citenamefont {Chiu}, \citenamefont {Chou},\ and\ \citenamefont
  {Schnyder}}]{chan20163}%
  \BibitemOpen
  \bibfield  {author} {\bibinfo {author} {\bibfnamefont {Y.-H.}\ \bibnamefont
  {Chan}}, \bibinfo {author} {\bibfnamefont {C.-K.}\ \bibnamefont {Chiu}},
  \bibinfo {author} {\bibfnamefont {M.~Y.}\ \bibnamefont {Chou}}, \ and\
  \bibinfo {author} {\bibfnamefont {A.~P.}\ \bibnamefont {Schnyder}},\ }\href
  {\doibase 10.1103/PhysRevB.93.205132} {\bibfield  {journal} {\bibinfo
  {journal} {Phys. Rev. B}\ }\textbf {\bibinfo {volume} {93}},\ \bibinfo
  {pages} {205132} (\bibinfo {year} {2016})}\BibitemShut {NoStop}%
\bibitem [{\citenamefont {Fang}\ \emph {et~al.}(2016)\citenamefont {Fang},
  \citenamefont {Weng}, \citenamefont {Dai},\ and\ \citenamefont
  {Fang}}]{fang2016topological}%
  \BibitemOpen
  \bibfield  {author} {\bibinfo {author} {\bibfnamefont {C.}~\bibnamefont
  {Fang}}, \bibinfo {author} {\bibfnamefont {H.}~\bibnamefont {Weng}}, \bibinfo
  {author} {\bibfnamefont {X.}~\bibnamefont {Dai}}, \ and\ \bibinfo {author}
  {\bibfnamefont {Z.}~\bibnamefont {Fang}},\ }\href {\doibase
  10.1088/1674-1056/25/11/117106} {\bibfield  {journal} {\bibinfo  {journal}
  {Chinese Physics B}\ }\textbf {\bibinfo {volume} {25}},\ \bibinfo {pages}
  {117106} (\bibinfo {year} {2016})}\BibitemShut {NoStop}%
\bibitem [{\citenamefont {Zak}(1989)}]{PhysRevLett.62.2747}%
  \BibitemOpen
  \bibfield  {author} {\bibinfo {author} {\bibfnamefont {J.}~\bibnamefont
  {Zak}},\ }\href {\doibase 10.1103/PhysRevLett.62.2747} {\bibfield  {journal}
  {\bibinfo  {journal} {Phys. Rev. Lett.}\ }\textbf {\bibinfo {volume} {62}},\
  \bibinfo {pages} {2747} (\bibinfo {year} {1989})}\BibitemShut {NoStop}%
\bibitem [{\citenamefont {Vanderbilt}\ and\ \citenamefont
  {King-Smith}(1993)}]{PhysRevB.48.4442}%
  \BibitemOpen
  \bibfield  {author} {\bibinfo {author} {\bibfnamefont {D.}~\bibnamefont
  {Vanderbilt}}\ and\ \bibinfo {author} {\bibfnamefont {R.~D.}\ \bibnamefont
  {King-Smith}},\ }\href {\doibase 10.1103/PhysRevB.48.4442} {\bibfield
  {journal} {\bibinfo  {journal} {Phys. Rev. B}\ }\textbf {\bibinfo {volume}
  {48}},\ \bibinfo {pages} {4442} (\bibinfo {year} {1993})}\BibitemShut
  {NoStop}%
\bibitem [{\citenamefont {Hirayama}\ \emph {et~al.}(2017)\citenamefont
  {Hirayama}, \citenamefont {Okugawa}, \citenamefont {Miyake},\ and\
  \citenamefont {Murakami}}]{hirayama2017topological}%
  \BibitemOpen
  \bibfield  {author} {\bibinfo {author} {\bibfnamefont {M.}~\bibnamefont
  {Hirayama}}, \bibinfo {author} {\bibfnamefont {R.}~\bibnamefont {Okugawa}},
  \bibinfo {author} {\bibfnamefont {T.}~\bibnamefont {Miyake}}, \ and\ \bibinfo
  {author} {\bibfnamefont {S.}~\bibnamefont {Murakami}},\ }\href
  {https://www.nature.com/articles/ncomms14022#citeas} {\bibfield  {journal}
  {\bibinfo  {journal} {Nature communications}\ }\textbf {\bibinfo {volume}
  {8}},\ \bibinfo {pages} {1} (\bibinfo {year} {2017})}\BibitemShut {NoStop}%
\bibitem [{\citenamefont {Liu}\ \emph {et~al.}(2020{\natexlab{b}})\citenamefont
  {Liu}, \citenamefont {Wang}, \citenamefont {Liu}, \citenamefont {Choi},
  \citenamefont {Kim}, \citenamefont {Jia}, \citenamefont {Park},\ and\
  \citenamefont {Cho}}]{liu2020ferromagnetic}%
  \BibitemOpen
  \bibfield  {author} {\bibinfo {author} {\bibfnamefont {S.}~\bibnamefont
  {Liu}}, \bibinfo {author} {\bibfnamefont {C.}~\bibnamefont {Wang}}, \bibinfo
  {author} {\bibfnamefont {L.}~\bibnamefont {Liu}}, \bibinfo {author}
  {\bibfnamefont {J.-H.}\ \bibnamefont {Choi}}, \bibinfo {author}
  {\bibfnamefont {H.-J.}\ \bibnamefont {Kim}}, \bibinfo {author} {\bibfnamefont
  {Y.}~\bibnamefont {Jia}}, \bibinfo {author} {\bibfnamefont {C.~H.}\
  \bibnamefont {Park}}, \ and\ \bibinfo {author} {\bibfnamefont {J.-H.}\
  \bibnamefont {Cho}},\ }\href {https://arxiv.org/abs/2007.05695} {\bibfield
  {journal} {\bibinfo  {journal} {arXiv preprint arXiv:2007.05695}\ } (\bibinfo
  {year} {2020}{\natexlab{b}})}\BibitemShut {NoStop}%
\bibitem [{\citenamefont {Mattausch}\ \emph {et~al.}(2005)\citenamefont
  {Mattausch}, \citenamefont {Zheng}, \citenamefont {Ryazanov},\ and\
  \citenamefont {Simon}}]{mattausch2005reduced}%
  \BibitemOpen
  \bibfield  {author} {\bibinfo {author} {\bibfnamefont {H.}~\bibnamefont
  {Mattausch}}, \bibinfo {author} {\bibfnamefont {C.}~\bibnamefont {Zheng}},
  \bibinfo {author} {\bibfnamefont {M.}~\bibnamefont {Ryazanov}}, \ and\
  \bibinfo {author} {\bibfnamefont {A.}~\bibnamefont {Simon}},\ }\href
  {\doibase 10.1002/zaac.200400269} {\bibfield  {journal} {\bibinfo  {journal}
  {Zeitschrift für anorganische und allgemeine Chemie}\ }\textbf {\bibinfo
  {volume} {631}},\ \bibinfo {pages} {302} (\bibinfo {year}
  {2005})}\BibitemShut {NoStop}%
\bibitem [{\citenamefont {Chiu}\ and\ \citenamefont
  {Schnyder}(2014)}]{PhysRevB.90.205136}%
  \BibitemOpen
  \bibfield  {author} {\bibinfo {author} {\bibfnamefont {C.-K.}\ \bibnamefont
  {Chiu}}\ and\ \bibinfo {author} {\bibfnamefont {A.~P.}\ \bibnamefont
  {Schnyder}},\ }\href {\doibase 10.1103/PhysRevB.90.205136} {\bibfield
  {journal} {\bibinfo  {journal} {Phys. Rev. B}\ }\textbf {\bibinfo {volume}
  {90}},\ \bibinfo {pages} {205136} (\bibinfo {year} {2014})}\BibitemShut
  {NoStop}%
\bibitem [{\citenamefont {Hirayama}\ \emph
  {et~al.}(2018{\natexlab{b}})\citenamefont {Hirayama}, \citenamefont
  {Okugawa},\ and\ \citenamefont {Murakami}}]{doi:10.7566/JPSJ.87.041002}%
  \BibitemOpen
  \bibfield  {author} {\bibinfo {author} {\bibfnamefont {M.}~\bibnamefont
  {Hirayama}}, \bibinfo {author} {\bibfnamefont {R.}~\bibnamefont {Okugawa}}, \
  and\ \bibinfo {author} {\bibfnamefont {S.}~\bibnamefont {Murakami}},\ }\href
  {\doibase 10.7566/JPSJ.87.041002} {\bibfield  {journal} {\bibinfo  {journal}
  {Journal of the Physical Society of Japan}\ }\textbf {\bibinfo {volume}
  {87}},\ \bibinfo {pages} {041002} (\bibinfo {year}
  {2018}{\natexlab{b}})}\BibitemShut {NoStop}%
\bibitem [{\citenamefont {Setyawan}\ and\ \citenamefont
  {Curtarolo}(2010{\natexlab{b}})}]{setyawan2010high-throughput}%
  \BibitemOpen
  \bibfield  {author} {\bibinfo {author} {\bibfnamefont {W.}~\bibnamefont
  {Setyawan}}\ and\ \bibinfo {author} {\bibfnamefont {S.}~\bibnamefont
  {Curtarolo}},\ }\href {\doibase
  https://doi.org/10.1016/j.commatsci.2010.05.010} {\bibfield  {journal}
  {\bibinfo  {journal} {Computational Materials Science}\ }\textbf {\bibinfo
  {volume} {49}},\ \bibinfo {pages} {299 } (\bibinfo {year}
  {2010}{\natexlab{b}})}\BibitemShut {NoStop}%
\end{thebibliography}%
	
\end{document}